\documentclass[conference, final]{IEEEtran}

\IEEEoverridecommandlockouts
\usepackage{hyperref}
\usepackage{tabularx}
\usepackage[nolist]{acronym}
\usepackage{listings}
\usepackage{amsmath,amssymb,amsfonts}
\usepackage{algorithmic}
\usepackage{graphicx}
\usepackage{textcomp}
\usepackage{xcolor}
\usepackage{float}
\usepackage{placeins}
\usepackage{templates/ebpackages}
\usepackage{todonotes}
\usepackage{tikz,pgfplots}
\usepackage{cleveref}
\usepackage{verbatim}
\usepackage{booktabs}
\usepackage{caption}
\usepackage[super]{nth}
\usepackage{makecell}

\usepackage[citestyle=numeric,
	style=numeric,
	backend=biber,
	doi=false,url=false,
	isbn=false]{biblatex}	
\usepackage{hhline}
\begin{acronym}[WS-Agreement] 
\acro{AADL}{Architecture Analysis and Design Language \acroextra{(originally: Avionics Architecture Description Language)}}
\acro{ADL}{architecture description language}
\acro{DPE}{Declarative Performance Engineering}
\acro{ADM}{Architecture-Driven Modernization}
\acro{ADML}{Architecture Description Markup Language}
\acro{AMI}{Amazon Machine Image}
\acro{AML}{architecture modification language}
\acro{AMQP}{Advanced Message Queuing Protocol}
\acro{AOP}{aspect-oriented programming}
\acro{API}{application programming interface}
\acro{APM}{Application Performance Management}
\acro{ARIMA}{autoregressive integrated moving average}
\acro{ARM}{Application Response Management}
\acro{AS}{Application Server}
\acro{ASP}{application service provider}
\acro{ASTM}{Abstract Syntax Tree Metamodel}
\acro{ATL}{ATLAS Transformation Language}
\acro{AWS}{Amazon Web Services}
\acro{Bash}{Bourne-again shell}
\acro{BMBF}{German Federal Ministry of Education and Research}
\acro{BNF}{Backus–Naur Form}
\acro{BPMN2}{Business Process Model and Notation, version~2.0}
\acro{CART}{Classification And Regression Tree} 
\acro{CBMG}{Customer Behavior Model Graph}
\acro{CBSA}{component-based software architecture}
\acro{CBSE}{component-based software engineering}
\acro{CBSPE}[CB-SPE]{Component-Based Software Performance Engineering}
\acro{CBSS}{component-based software system}
\acro{CCM}{\acs{CORBA} Component Model}
\acro{CD}{continuous deployment}
\acro{CDF}{Cumulative Distribution Function}
\acro{CDFs}{Cumulative Distribution Functions}
\acro{CDN}{content delivery network}
\acro{CDO}{Connected Data Objects}
\acro{CEP}{complex event processing}
\acro{CEST}{Central European Summer Time}
\acro{CET}{Central European Time}
\acro{CI}{continuous integration}
\acro{CIM}{Common Information Model}
\acro{CLI}{command-line interface}
\acro{CMOF}{Complete \acs{MOF}}
\acro{COBOL}{Common Business-Oriented Language}
\acro{CoCoME}{Common Component Modeling Example}
\acro{CIPM}{Continuous Integration of the Performance Model}
\acro{PMPs}{Performance Model Parameters}
\acro{PM}{Performance Model}
\acro{PMs}{Performance Models}
\acro{COM}{Component Object Model}
\acro{CORBA}{Common Object Request Broker Architecture}
\acro{CPN}{Colored Petri Net}
\acro{CPU}{central processing unit}
\acro{CRM}{customer relationship management}
\acro{CRUD}{create, read, update, and delete}
\acro{CSE}{Continuous Software Engineering}
\acro{CSM}{Core Scenario Model}
\acro{CSV}{comma-separated values}
\acro{CTMC}{continuous-time Markov chain}
\acro{CM} {Correspondence Model}
\acro{DBMS}{database management system}
\acro {Dev-changes}{Development changes}
\acro{DCOM}{Distributed Component Object Model}
\acro{Dev-changes}{development changes}
\acro{Dev-time}{Development time}
\acro{Ops-time}{Operation time}
\acro{DFG}{Deutsche Forschungsgemeinschaft (German Research Foundation)}
\acro{DML}{Descartes Modeling Language}
\acro{DNS}{Domain Name System}
\acro{DSL}{domain-specific language}
\acro{DTMC}{discrete-time Markov chain}
\acro{DQL}{Descartes Query Language}
\acro{EAS}{enterprise application system}
\acro{EBS}{Elastic Block Store}
\acro{EC2}{Elastic Compute Cloud}
\acro{Eclipse}{Eclipse \acs{IDE} and \acs{RCP}}
\acro{EJB}{Enterprise JavaBeans}
\acro{EMF}{Eclipse Modeling Framework}
\acro{EMI}{Eucalyptus Machine Image}
\acro{EMOF}{Essential \acs{MOF}}
\acro{EMP}{Eclipse Modeling Project}
\acro{EPL}{Event Processing Language}
\acro{EQN}{Extended Queueing Network}
\acro{ERP}{enterprise resource planning}
\acro{EU}{European Union}
\acro{FP7}{Seventh Framework Programme by the \acs{EU}}
\acro{GCS}{graphical concrete syntax}
\acro{GPL}{general purpose language}
\acro{GQM}{Goal Question Metric}
\acro{GSLA}{Generalized Service Level Agreement}
\acro{GXLA}{\acs{XML}-based implementation of \acs{GSLA}}
\acro{HDD}{hard disk drive}
\acro{HOT}{higher order transformation}
\acro{HSD}{Honestly Significant Difference}
\acro{HTML}{HyperText Markup Language}
\acro{HTTP}{Hypertext Transfer Protocol}
\acro{HUTN}{Human-Usable Textual Notation}
\acro{IaaS}{infrastructure as a service}
\acro{ICAC}{International Conference on Autonomic Computing}
\acro{ICPE}{International Conference on Performance Engineering}
\acro{ICT}{information and communication technology}
\acro{IDE}{integrated development environment}
\acro{IEC}{International Electrotechnical Commission}
\acro{IEEE}{Institute of Electrical and Electronics Engineers}
\acro{IMM}{Instrumentation MetaModel}
\acro{IM}{Instrumentation Model}
\acro{IP}{Internet Protocol}
\acro{IRL}{Instrumentation Record Language}
\acro{ISO}{International Organization for Standardization}
\acro{IT}{information technology}
\acro{J2EE}{Java 2 Platform, Enterprise Edition}
\acro{JaMoPP}{Java Model Parser and Printer}
\acro{Java}{Java}
\acro{JavaEE}[Java\,EE]{\acs{Java} Platform, Enterprise Edition}
\acro{JavaSE}[Java\,SE]{Java Platform, Standard Edition}
\acro{JCP}{\acs{Java} Community Process}
\acro{JDBC}{\acs{Java} Database Connectivity} 
\acro{JMS}{\acs{Java} Message Service}
\acro{JMT}{\acs{Java} Modeling Tools}
\acro{JMX}{\acs{Java} Management Extensions}
\acro{JNDI}{\acs{Java} Naming and Directory Interface}
\acro{JPA}{\acs{Java} Java Persistence \acs{API}}
\acro{JRE}{\acs{Java} Runtime Environment}
\acro{JSF}{JavaServer Faces}
\acro{JSP}{JavaServer Pages}
\acro{JSR}{\acs{Java} Specification Request}
\acro{JTA}{\acs{Java} Transaction \acs{API}}
\acro{JVM}{\acs{Java} Virtual Machine}
\acro{JVMPI}{\acs{JVM} Profiler Interface}
\acro{JVMTI}{\acs{JVM} Tool Interface}
\acro{KDM}{Knowledge Discovery Meta-Model}
\acro{KIT}{Karlsruhe Institute of Technology}
\acro{KLOC}{\acsp{LOC} in thousands}
\acro{KM3}{Kernel Meta Meta Model}
\acro{KoSSE}{Kompetenzverbund Software Systems Engineering}
\acro{KPI}{Key Performance Indicator}
\acro{KS-Test}{Kolmogorov-Smirnov-Test}
\acro{LAN}{local area network}
\acro{LibReDE}{Library for Resource Demand Estimation}
\acro{LOC}{lines of code}
\acro{LQN}{Layered Queueing Network}
\acro{M2M}{model-to-model}
\acro{M2T}{model-to-text}
\acro{MAMBA}{Measurement Architecture for Model-Based Analysis}
\acro{MAPE}{Mean Absolute Percentage Error}
\acro{MAPEK}[MAPE-K]{Modeling, Analysis, Planning, Execution, Knowledge}
\acro{MARTE}{\acs{UML} Profile for MARTE: Modeling and Analysis of Real-Time Embedded Systems}
\acro{AbPP}{Architecture-based Performance Prediction}
\acro{MbDevOps}{Model-based DevOps} 
\acro{MDA}{Model-Driven Architecture}
\acro{MDD}{model-driven development}
\acro{MDE}{model-driven engineering}
\acro{MDSD}{model-driven software development}
\acro{MDSE}{model-driven software en\-gi\-neer\-ing}
\acro{MIAs}{Monitored Internal Actions}
\acro{MMM}{Measurements MetaModel}
\acro{MM}{Measurements Model}
\acro{MOF}{Meta Object Facility}
\acro{MOM}{message-oriented middleware}
\acro{MTBF}{mean time between failure}
\acro{MTTF}{mean time to failure}
\acro{MTTR}{mean time to repair}
\acro{MVA}{Mean Value Analysis}
\acro{NATO}{North Atlantic Treaty Organization}
\acro{NIST}{National Institute of Standards and Technology}
\acro{NMIAs}{Not Monitored Internal Actions}
\acro{NTP}{Network Time Protocol}
\acro{oAW}{openArchitectureWare}
\acro{OCL}{Object Constraint Language}
\acro{OGF}{Open Grid Forum}
\acro{OMG}{Object Management Group}
\acro{OPAD}[$\Theta$PAD]{Online Performance Anomaly Detection}
\acro{Ops-changes}{operation changes}
\acro{OSGi}{A \acs{Java} component model (originally, \acs{OSGi} was an acronym for  Open Services Gateway initiative)}
\acro{OZ}[Object-Z]{Object-Z Specification Language}
\acro{PaaS}{platform as a service}
\acro{PCA}{Power Consumption Analyzer}
\acro{PCM}{Palladio Component Model}
\acro{PMX}{Performance Model Extractor}
\acro{PN}{Petri Net}
\acro{PMIF}{Performance Model Interchange Format}
\acro{QFTP}{\acs{UML} Profile for Modeling Quality of Service and Fault Tolerance Characteristics and Mechanisms}
\acro{QM}{Queueing Model}
\acro{QN}{Queueing Network}
\acro{QPME}{Quereing Petri Net Modeling Environment}
\acro{QoS}[QoS]{quality of service}
\acro{QPN}{Queueing Petri Net}
\acro{QVT}{Query/View/Transformation}
\acro{RAC}{Run-time Architecture Correspondence Meta-Model}
\acro{RCP}{rich client platform}
\acro{RDS}{Relational Database Service}
\acro{RD}{Resource Demand}
\acro{RDs}{Resource Demands}
\acro{RDE}{Resource Demand Estimation}
\acro{RDSEFF}{Resource Demanding \acs{SEFF}}
\acro{REST}{Representational State Transfer}
\acro{RMI}{Remote Method Invocation}
\acro{RMSE}{Root Mean Squared Error}
\acro{RPC}{remote procedure call}
\acro{RRA}{Round Robin Archive}
\acro{RRD}{Round Robin Database}
\acro{RRDtool}{Round Robin Database tool} 
\acro{RSA}{Rational Software Architect}
\acro{S3}{Simple Storage Service}
\acro{SaaS}{software as a service}
\acro{SAMM}{Service Architecture Meta Model}
\acro{SASS}{self-adaptive software system}
\acro{SCP}{Secure Copy}
\acro{SEAMS}{Symposium on Software Engineering for Adaptive and Self-Managing Systems}
\acro{SEFF}{Service Effect Specification}
\acro{SEFFs}{Service Effect Specifications} 
\acro{SEI}{Carnegie Mellon Software Engineering Institute}
\acro{SLA}{service level agreement}
\acro{SLAstar}[SLA$\star$]{a language for specifying \acsp{SLA}}
\acro{SLAng}{a language for specifying \acsp{SLA}}
\acro{SLAstic}{name of the approach developed in this thesis}
\acro{SLM}{service level management}
\acro{SLO}{service level objective}
\acro{SLS}{service level specification}
\acro{SMM}{Structured Metrics Meta-Model}
\acro{SMML}{Software Measurement Modeling Language}
\acro{SOA}{service-oriented architecture}
\acro{SOAP}{protocol for exchanging structured information in computer networks (originally: Simple Object Access Protocol)}
\acro{SoMoX}{Software Model eXtractor}
\acro{SPE}{Software Performance Engineering}
\acro{SPEC}{Standard Performance Evaluation Corporation}
\acro{SPECRG}[SPEC RG]{\acs{SPEC} Research Group}
\acro{SPEL}{Software Performance Engineering Lab}
\acro{SPN}{Stochastic Petri Nets}
\acro{SPPDFF}[SPP~1593]{Priority Programme ``Design for Future --- Managed Software Evolution''}
\acro{SPT}{\acs{UML} Profile for Schedulability, Performance, and Time}
\acro{StoEx} {stochastic expression}
\acro{S/T/A}{Strategies/Tactics/Actions \acroextra{(modeling language)}}
\acro{SQL}{Structured Query Language}
\acro{SUT}{system under test}
\acro{SVM}{Support Vector Machine}
\acro{SSH}{Secure Shell}
\acro{SVN}{Subversion}
\acro{SWaP}{Space, Watts and Performance}
\acro{T2M}{text-to-model}
\acro{TCO}{total cost of ownership}
\acro{TCPIP}[TCP/IP]{Transmission Control Protocol/Internet Protocol}
\acro{TCS}{textual concrete syntax}
\acro{TPN}{Timed Petri Net}
\acro{UC}{University of California}
\acro{UCM}{Use Case Map}
\acro{UI}{user interface}
\acro{UML}{Unified Modeling Language}
\acro{UML2}[UML\,2]{\acs{UML}, version~2}
\acro{UNISTUTT}[UNI-STUTT]{University of Stuttgart}
\acro{UNIWUERZ}[UNI-WUERZ]{University of Würzburg}
\acro{URI}{Uniform Resource Identifier}
\acro{URL}{Uniform Resource Locator \acroextra{(originally: Universal Resource Locator)}}
\acro{URN}{Uniform Resource Name}
\acro{UTC}{Coordinated Universal Time}
\acro{VB6}{Visual Basic~6}
\acro{VCS}{version control system}
\acro{VSUM}{Virtual Single Underlying Model}
\acro{VSUMM}{Virtual Single Underlying MetaModel}
\acro{W3C}{World Wide Web Consortium}
\acro{WCOP}{Workshop on Component-Oriented Programming}
\acro{WS-Agreement}{Web Service Agreement}
\acro{WSDL}{Web Services Description Language}
\acro{WSLA}{Web Service Level Agreement}
\acro{WSOI}{Web Service Offerings Infrastructure}
\acro{WSOL}{Web Service Offerings Language}
\acro{XMI}{\acs{XML} Metadata Interchange}
\acro{XML}{eXtensible Markup Language}
\acro{Xtend}{a programming language on top of \acs{Xtext}}
\acro{Xtext}{framework for development of programming languages and \acsp{DSL}}
\acro{Z}{Z Specification Language}
\end{acronym}  
\begin{document}
\IEEEoverridecommandlockouts
\IEEEpubid{
\begin{minipage}{\textwidth}\ \\[12pt]{
 \textbf{978-1-7281-4659-1/20/\$31.00~\copyright2020 IEEE.}
\\
\textbf{DOI: 10.1109/ICSA47634.2020.00011}}
\end{minipage}
}

\title{Incremental Calibration of Architectural Performance Models with Parametric Dependencies}

\author{
\IEEEauthorblockN{Manar Mazkatli}
\IEEEauthorblockA{\textit{Karlsruhe Institute of }\\
\textit{Technology, Germany} \\
manar.mazkatli@kit.edu}
\and
\IEEEauthorblockN{David Monschein}
\IEEEauthorblockA{\textit{Karlsruhe Institute of }\\
\textit{Technology, Germany} \\
david.monschein@student.kit.edu}
\and
\IEEEauthorblockN{Johannes Grohmann}
\IEEEauthorblockA{{\textit{University of W{\"u}rzburg}} \\
\textit{W{\"u}rzburg, Germany}\\
 johannes.grohmann@uni-wuerzburg.de}
\and
\IEEEauthorblockN{Anne Koziolek}
\IEEEauthorblockA{\textit{Karlsruhe Institute of }\\
\textit{Technology, Germany} \\
koziolek@kit.edu}
}
\makeatletter
\patchcmd{\@maketitle}
  {\addvspace{0.5\baselineskip}\egroup}
  {\addvspace{-1\baselineskip}\egroup}
  {}
  {}
\makeatother
\maketitle

\begin{abstract}
 Architecture-based Performance Prediction (AbPP) allows evaluation of the performance of systems and to answer what-if questions without measurements for all alternatives. A difficulty when creating models is that Performance Model Parameters (PMPs, such as resource demands, loop iteration numbers and branch probabilities) depend on various influencing factors like input data, used hardware and the applied workload. To enable a broad range of what-if questions, Performance Models (PMs) need to have predictive power beyond what has been measured to calibrate the models. Thus, PMPs need to be parametrized over the influencing factors that may vary.  

Existing approaches allow for the estimation of parametrized PMPs by measuring the complete system. Thus, they are too costly to be applied frequently, up to after each code change. They do not keep also manual changes to the model when recalibrating.

In this work, we present the Continuous Integration of Performance Models (CIPM), which \emph{incrementally} extracts and calibrates the performance model, including parametric dependencies.
CIPM responds to source code changes by updating the PM and adaptively instrumenting the changed parts. To allow AbPP, CIPM estimates the parametrized PMPs using the measurements (generated by performance tests or executing the system in production) and statistical analysis, e.g., regression analysis and decision trees. 
Additionally, our approach responds to production changes (e.g., load or deployment changes) and calibrates the usage and deployment parts of PMs accordingly.

For the evaluation, we used two case studies. 
Evaluation results show that we were able to calibrate the PM incrementally and accurately.

\end{abstract}
\vspace{-0.1 cm}\section{Introduction}
\label{Introduction}
In many application domains, software is nowadays developed iteratively and incrementally, e.g., following agile practices.
This means that there usually is no dedicated architecture design phase, but architectural design decisions are made continuously throughout the development. For performance analysis, this means that there is not a single point in time at which the performance analysis is carried out, but a performance analysis is required whenever performance-critical architecture design decisions have to be made. 

In such iterative and incremental development projects, developers typically use \ac{APM} \cite{apm,BeEiFeGrRhJaShHoViWaWi2019-ICPE-DevOpsSurvey} to assess the current performance of their software. However, \ac{APM} cannot predict the impact of design decisions or to answer scalability and sizing questions in a large and distributed environment, where the resources may not be accessible and a huge amount of load generators would be needed. 

Software  performance  engineering \cite{SPEwoodside,conni} uses models to assess the performance of a software system. It supports the evaluation of various architecture, design, implementation and workload choices. Compared to \ac{APM}, such model-based performance prediction \cite{mbpp} allows to predict future alternatives before implementing them \cite{conni}.
In particular, \ac{AbPP} approaches~\cite{reussner2016a} reflect the architecture of the software system in a \ac{PM} in order to easily study design alternatives~\cite{martens2010c}. 

The problem of \ac{AbPP} in iterative-incremental development is that modelling is a time-consuming process. In particular, calibrating \ac{PMPs} (e.g., \ac{RDs}, branch probabilities, and loop counts) is challenging, because the \ac{PMPs} may depend on impacting factors such as input data, properties of required resources or usage profile. Ignoring these so-called \emph{parametric dependencies}~\cite{becker2008a}) will lead to inaccurate performance predictions. Thus, keeping the \ac{PM} and in particular the parametrized \ac{PMPs} consistent with the iteratively and incrementally evolving source code over time requires repeated manual effort.

To address the high effort to create performance models when source code is available, researchers have suggested several approaches to extract \acp{PM} automatically.
However, most of these approaches \cite{BrHuKo2011-ASE-AutomExtraction,WaStKoKo2017-QUDOS-PMXBuilder, Brunnert13,langhammer2016automated,spinner2016reference} have three shortcomings: First, they require instrumentation and execution  of the whole system under study to extract the \ac{PM}, which causes a high overhead and is not feasible at high frequency, e.g., after each source code commit. Second, they do not consider parametric dependencies. 
Notable exceptions are the approach by Krogmann et al.~\cite{krogmann2009ck,krogmann2012ashort} or Grohmann et al.~\cite{GrEiElMaKiKo2019-MASCOTS-DependencyIdentification}, which also extract parametric dependencies (but monitor the whole system to calibrate the model without keeping the manual changes). 
Third, they reconstruct the whole architectural PM from scratch. Thus, they cannot keep manual changes of the architecture model, which is problematic if the architecture model shall be used as an architecture knowledge base and is enhanced manually with e.g. the rationale of architecture decisions~\cite{kramer2012b}. 

In this paper, we present our approach, called \ac{CIPM}, to incrementally extract and calibrate architecture-level \acp{PM} with parametric dependencies after each source code commit. 
Our approach builds upon the Palladio approach~\cite{reussner2016a} for modelling and simulating architecture-level \acp{PM}. We have extended an incremental extraction of such architectural \acp{PM}~\cite{langhammer2017a} with incremental calibration that considers the parametric dependencies and is based on adaptive monitoring. The goal of the approach is to keep the \ac{PM} up-to-date automatically to allow \ac{AbPP}. An initial version of this idea has been presented in a workshop publication~\cite{Mazkatli2018Qudos} without evaluation.

The contributions of our paper are twofold: 
\begin{enumerate} 
\renewcommand{\labelenumi}{(\Alph{enumi})}
\item \textit{Incremental Dev-time calibration:} We propose a novel incremental calibration at \ac{Dev-time}  that responds to source code changes by adaptive instrumentation of the changed parts of the code and uses the resulting measurements from performance tests or the production system to estimate the \ac{PMPs} incrementally.

For this purpose, we propose a novel incremental \ac{RDE} that is based on adaptive monitoring.
Our calibration uses statistical analysis to learn potential dependencies, e.g., regression analysis for resource demands and decision trees for the estimation of  branch transitions.

\item \textit{Model-based DevOps pipeline:}
We extend agile DevOps practices \cite{brunnert2015a}  to integrate the \ac{CIPM} activities and thus to reduce the effort for \ac{AbPP}. 
To do so, we propose a model-based DevOps pipeline and implement a part of it. 
In addition to the above-mentioned incremental \ac{Dev-time} calibration, we also include an existing approach for calibration at \ac{Ops-time} in the pipeline. This \ac{Ops-time} calibration responds to changes in deployment and usage profile and updates the respective parts of the \ac{PM}. We include also prototypical self-validation steps.



\end{enumerate}
We evaluate our contributions using two case studies \cite{cocome,teastore}. 
The evaluation confirmed that the incremental calibration was able to detect parametric dependencies while significantly reducing the monitoring overhead. Moreover, we showed that the calibrated performance models are accurate by comparing the simulation results with monitored data.

The next section gives an overview of the foundations. \Cref{Running Example} introduces a motivating example. The overall CIPM process is presented in \Cref{CIPM}. \Cref{MbDevOps} describes how to embed CIPM in the model-based DevOps pipeline. The next three sections provide detail on the CIPM activities: \Cref{adaptive instrumentation} presents the adaptive instrumentation, \Cref{incremental calibration} describes the incremental calibration and \Cref{parametric dependencies} describes how we estimate the parametric dependencies. Our evaluations are discussed in \Cref{evaluation}. The related work follows in \Cref{related works}. The paper ends with our conclusions and future work 
(\Cref{Conclusion}).
 
\vspace{-0.1 cm}
\section{Foundations} 
\label{foundation}
\subsection{Palladio}
\label{palladio}
Palladio is an approach to model and simulate architecture-level \acp{PM} and has been used in various industrial projects~\cite{reussner2016a,becker2008a}. Within Palladio, the so-called \ac{PCM} defines a language for describing \acp{PM}: the static structure of the software (e.g. components and interfaces), the behavior, the required resource environment, the allocation of software components and the usage profile. The \ac{PCM} \ac{SEFF} \cite{becker2008a} describes the behavior  of a component service on an abstract level using different control flow elements: \textit{internal actions} (a combination of internal computations that do not include calls to required services), \textit{external call actions} (calls to required services), \textit{loops} and \textit{branch actions}. SEFF loops and branch actions include at least one external call. Other loops and branches in the service implementation are ignored and combined into the  internal actions to increase the level of abstraction.

To predict the performance measures (response times, \ac{CPU} utilization and throughput) the architects  have to enrich the SEFFs with \ac{PMPs}.
Examples of \ac{PMPs} are resource demands (processing amount that internal action requests from a certain active resource, such as a \ac{CPU} or hard disk), the probability of selecting a branch, the number of loop iterations and the arguments of external calls.

Palladio uses the \ac{StoEx} language \cite{koziolek_modeling_2016} to define \ac{PMPs} as expressions that contain random variables or empirical distributions.
\ac{StoEx} allows parameter characterization (e.g. determining NUMBER\_OF\_ELEMENTS, VALUE, BYTESIZE and TYPE) and to express calculations and comparisons (e.g., file.BYTESIZE$<=$5*max.VALUE).

\begin{figure*} 
	\centering
    \includegraphics[width=0.9\textwidth]{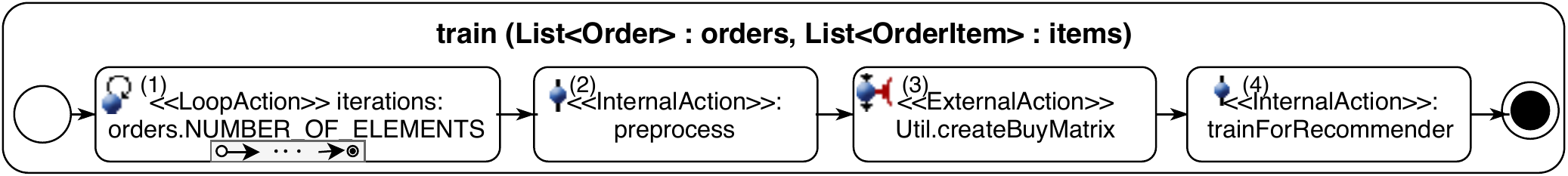}  
    \vspace{-0.2 cm}
    \caption{The behavior of TeaStore's "Recommender.train" service using SEFF}
    \label{fig:seff:train}
    \vspace{-0.3 cm}
\end{figure*}
\vspace{-0.07 cm}

\subsection{Co-evolution approach with \textsc{\textit{Vitruvius}}\xspace}
\label{vitruv}
\label{co-evaluation approach}
 The co-evolution approach of Langhammer et al. \cite{langhammer2015a,langhammer2017a}   helps software developers and architects to keep the architecture model and the code consistent when their software system evolves. It defines change-driven consistency preservation rules that propagate changes in source code to the architecture model and vice versa using model-based transformations.

These rules are defined based on \vitruv  \cite{kramer2013b,burger2014diss}, a view-based framework that encapsulates the heterogeneous models of a system and the semantic relationships between them in a \ac{VSUM} in order to keep them consistent. 
\vitruv defines \emph{mappings} and \emph{reactions} languages that describe consistency rules on the metamodel-level. These rules describe the consistency repair logic for each kind of changes, i.e., which and how the artifacts of a metamodel have to be changed to restore the consistency after a change in a related metamodel has occurred. 

Using \vitruv, Langhammer et al. implemented their approach to keep Java source code (using an intermediate model \cite{heidenreich2010a})  and a \ac{PCM} model consistent. The defined consistency rules update the structure of a \ac{PCM} model  (i.e. components and interfaces) and its behavior  (in term of \acsp{SEFF}, but \textit{without} \ac{PMPs}) as a reaction to changes in the source code. Similarly, changes in PCM model are propagated to the Java source code. 

\subsection{Kieker}
\label{kieker}
Kieker \cite{kieker} is an APM tool that captures and analyzes execution traces from distributed software systems. It allows one to describe specific probes (data structures of the monitored information) using the \ac{IRL}  \cite{jung2013model}. It also supports dynamic and adaptive monitoring by activating or deactivating probes during run time.

\subsection{iObserve}
\label{iObserve}

iObserve \cite{heinrich2017a} aims to increase the human comprehensibility of run-time changes by updating an architecture model accordingly.  The authors defined specific monitoring records using \ac{IRL} and map the resulting measurements to the corresponding parts in the \ac{PCM} using the so-called \ac{RAC} \cite{heinrich2014}. iObserve detects changes concerning the deployment and the user behavior  and updates the related parts of the \ac{PCM} to analyze performance and privacy aspects \cite{heinrich2016b}. However, iObserve does not support updating component behavior  models (SEFFs) with \ac{PMPs}.

\vspace{-0.1 cm}
\section{Running Example}
\label{Running Example}
We introduce a motivating example that illustrates our approach and was used to evaluate it (see Sec. \ref{evaluation}). The example is part of the TeaStore case study \cite{teastore}: a website to buy different kinds of tea. In this case study, the Recommender component is responsible for calculating the recommendations for a certain shopping cart using the services \textit{‘train’} and \textit{‘recommend’}. The \textit{‘train’} service derives information from the previous orders and prepares the data for the \textit{‘recommend’} service. Because there are different strategies to recommend a list of related items, the developers implemented four versions of \textit{‘recommend’}  and \textit{'train'} along different development iterations. 
These implementations have different performance characteristics. Performance tests or monitoring can be used to discover these characteristics for the current state, i.e., for the current deployment and the current workload. However, predicting the performance for another state (e.g., different deployment or workload) is  expensive and challenging because it requires setting up and performing several tests for each implementation alternative. 
In our example, answering the following questions is challenging based on APM: “Which implementation would perform better if the load or the deployment is changed?” or “How well does the \textit{'train'} service perform during yet unseen workload scenarios?” An example for the latter question would be an upcoming offer of discounts, where architects expect an increased number of customers and also a changed behavior  of customers in that each customer is expected to order more items.

AbPP can answer these questions faster using simulations instead of the expensive tests if an up-to-date PM is available.

Regardless how the model will be built and updated (reverse engineering extraction or manual/ automatic update), all available approaches recalibrate the whole model by monitoring all parts of the source code instead of recalibrating only the model parts affected by the last changes in source code. For example, the changes in the implementation of \textit{'train'}  belong to the last part of the code which is represented as an internal action ‘trainForRecommender’ by modelling the behaviour using SEFF (see Fig. \ref{fig:seff:train}). This means that these changes have only impact on the RD of ‘trainForRecommender’ and all other PMPs are valid (e.g., RD of preprocessor internal action). Recalibrating the whole PMPs loses potential previous manual changes and causes unnecessary monitoring overhead. Additionally, not all the available approaches detect the parametric dependencies (e.g., the RD of ‘trainForRecommender’ is related to the number of ordered items).


\begin{figure*}
\vspace{-0.2 cm}
	\centering
    \includegraphics[width=0.88\textwidth]{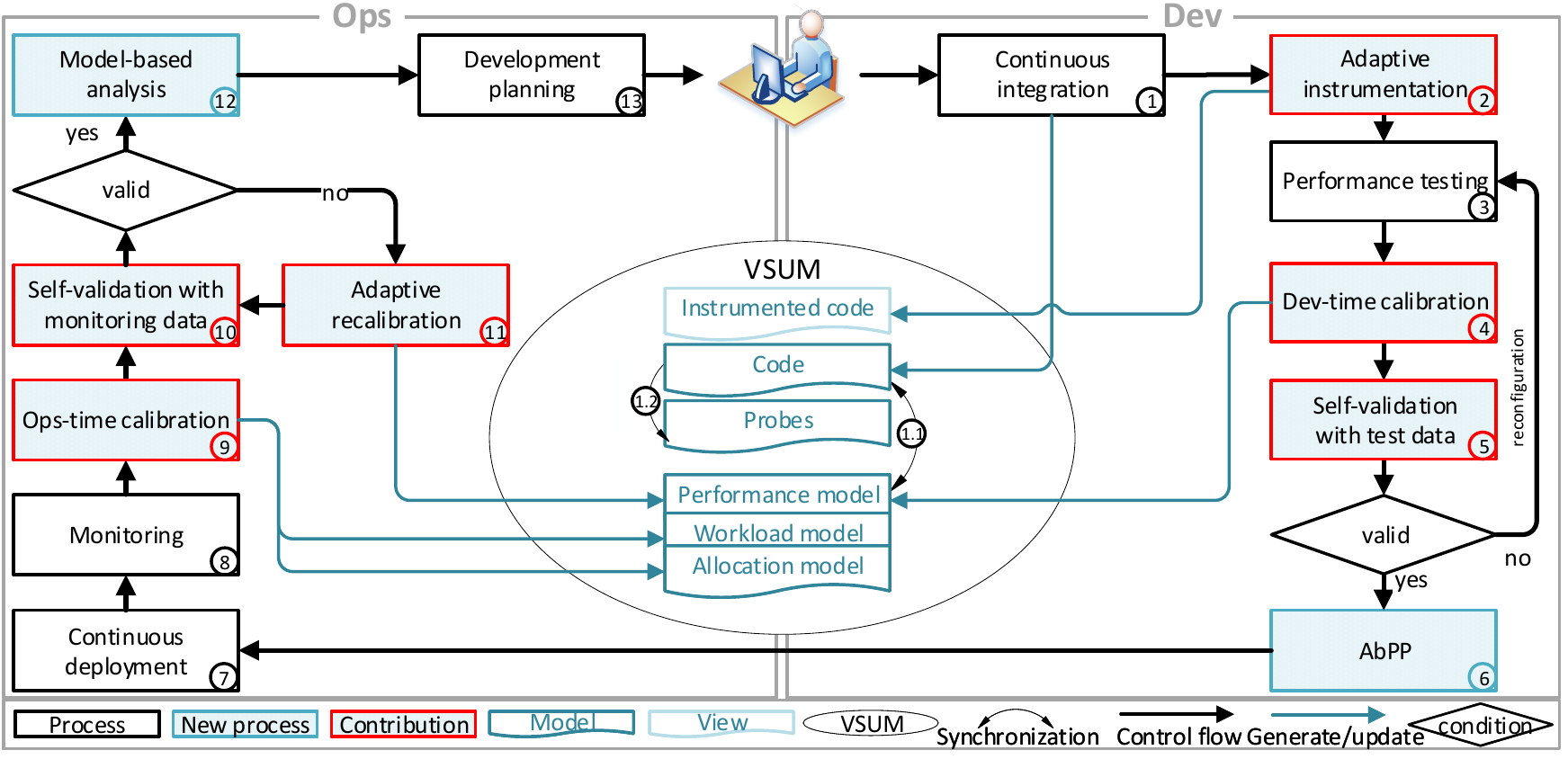}
    \vspace{-0.2 cm}
    \caption{Model-based DevOps pipeline}
    \vspace{-0.5 cm}
    \label{pipeline}
\end{figure*}
\vspace{-0.1 cm}
\section{Continuous Integration of Performance Model}
\label{CIPM}
This section provides an overview of the \ac{CIPM} approach, before describing how it is embedded in a continuous software engineering approach in \Cref{MbDevOps} and before providing more details on the CIPM activities in Sections VI-VII.

Our approach \ac{CIPM} incrementally extracts and calibrates archi\-tecture-level \acp{PM} with parametric dependencies after each source code commit. 
To do so, CIPM updates the PM continuously to keep it consistent with the running system, i.e., the deployed source code and the last measurements.

CIPM consists of four main activities:  
 \begin{enumerate}
     \item Performance  model update and adaptive instrumentation: CIPM analyzes the source code changes, updates the architectural PM and static behavior  model  based on the co-evolution approach \cite{langhammer2017a} and  instruments the changed parts of code to calibrate the new/ updated related parts of the architecture as will be discussed in \Cref{adaptive instrumentation}. 
     \item Monitoring: CIPM collects the required measurements either during testing or executing the system in production. 
     \item Incremental calibration of \ac{PM} (Sec. \ref{incremental calibration}):  CIPM performs the \textit{Dev-time calibration} of behavior, i.e., \ac{PMPs} (Sec. \ref{Dev-time}) considering the parametric dependencies (Sec. \ref{parametric dependencies}) and  the \textit{Ops-time calibration}, i.e., updating the deployment and usage parts of \ac{PM} (Sec. \ref{Run-time calibration}). 
	\item Self-validation: 
	CIPM automatically starts a simulation and calculates the variation between the simulation results and the monitoring data to show the estimation error before performing \ac{AbPP}. 
 \end{enumerate}
\label{Preserving the consistency}
To realize these four activities, CIPM extends the co-evolution approach  with a monitoring process to keep the PM consistent with the last up-to-date measurements. The static structure and behavior update of PM  is already provided by the co-evolution approach. For the adaptive instrumentation, we (a) extend the \ac{VSUM} with an \ac{IM} that describes and manages the instrumentation points and (b) define the consistency rules between the IM and source code. The rules define how \vitruv has to respond to the changes in code by adaptive instrumentation of the changed parts with predefined monitoring probes. For the monitoring, we (c) define a \ac{MM} that describes the resulting monitoring records.  These records belong to specific types that are responsible for collecting the necessary monitoring information to calibrate the SEFF elements. Finally, for the incremental calibration and estimation of parametric dependencies, we (d) define the consistency rules between MM and PM, which analyze the monitoring data, calibrate the PMPs and update the deployment and usage parts of PM.
 
With our approach, we can also handle overlapping commits that occur while the monitoring and calibration of a previous commit are still ongoing by managing multiple copies of the \ac{PM} and keeping track of which model instance belongs to which version of the source code.  

%
\vspace{-0.1 cm}
\section{The Model-based DevOps Pipeline}
\label{MbDevOps}
DevOps practices aim to close the gap between the development and operations and to integrate them into one reliable process.
We extend the DevOps practices to integrate and automate the \ac{CIPM} in a \ac{MbDevOps} pipeline. This enables \ac{AbPP} during DevOps-oriented development. The next paragraphs explain the  pipeline processes. The lower case numbers refer to the process's number in Fig. \ref{pipeline}.
The \ac{MbDevOps} pipeline starts on the ``development'' side with \(continuous\:integration_{1}\:(CI)\)   process \cite{ci1} that merges the source code changes of the developers (cf.~ Fig. \ref{pipeline}). \vitruv responds to the changes in source code with the first CIPM activity: \(updating \:the \:architectural\;PM_{1.1}\)  and \(generating\: the\: required\: instrumentation\: points\:in\:IM_{1.2}\)  (cf. Sec. \ref{co-evaluation approach}).
Next, \(adaptive\: instrumentation_{2}\) instruments the changed parts of source code using the instrumentation points from \ac{IM} (Sec. \ref{adaptive instrumentation}). The following process is the \(performance\:testing_{3}\), which integrates the second CIPM activity 'Monitoring' to generate the necessary measurements for calibration. The pipeline divides the measurements into training and validation set. Afterwards, the \(Dev{-}time\: calibration_{4}\) (the first part of \nth{3} CIPM activity) enriches the \ac{PM} with \ac{PMPs} using the training set. \Cref{incremental calibration} describes the incremental calibration process whereas \Cref{parametric dependencies} explains how we parametrize the \ac{PMPs} with the influencing input parameters. After the calibration, the pipeline starts  the \(self{-}validation\:with\:test\:data_{5}\) process (the \nth{4} CIPM activity), which uses the validation set to evaluate the calibration accuracy.
If the model is deemed accurate, developers can  use the resulting \ac{PM} to answer the performance questions using \(AbPP_{6}\). If not, they can change the test configuration to recalibrate \ac{PM}. Answering the performance questions using \ac{AbPP} instead of the test-based performance prediction reduces both  the effort and cost of setting up the test environment to perform this prediction.
 
The ``operations'' side of \Cref{pipeline} starts on the \(continuous\: deployment_{7}\) in the production environment. CIPM \(Monitoring_{8}\) in the production environment generates the required run-time measurements for the  next process: \(Ops{-}time\: calibration_{9}\) (the second part of the \nth{3} CIPM activity). The ops-time calibration calibrates and updates both usage model and deployment model (Sec. \ref{Run-time calibration}). Next, the \(self{-}validation\:with\:monitoring\:data_{10}\)     (the \nth{4} CIPM activity) validates the estimated \ac{PMPs}. If the \ac{PM} is not accurate enough, the \(adaptive\:recalibration_{11}\) process  (Sec. \ref{incremental calibration})  recalibrates the inaccurate model parts using monitoring data. 
Finally, the developers can perform more \(model-based\: analysis_{12}\) on the resulting model, e.g., model-based auto scaling. Additionally, having an up-to-date descriptive PM supports the \(development\) \(planning_{13}\). This is due to the advantages of models:  increasing the understandability of the current version, modelling and evaluating design alternatives and answering what-if questions. 
\vspace{-0.3 cm}
\section{Adaptive Instrumentation}
\label{adaptive instrumentation}
The goal is to instrument the parts of source code, which have been changed and their changes may affect the validity of related PMPs. In our running example, only the source code of 'trainForRecommender' is instrumented to provide the required measurements to update the RD, whereas there is no need to instrument the loop or the code of ‘preprocess’, because these parts of code are not changed, consequently the old estimations of the PMPs remain valid.

To automate the adaptive instrumentation,  we use the consistency rules between the source code model and IM, which responds to changes in method body (e.g., adding/ updating statements) by instrumenting the related parts. The rules reconstruct the \ac{SEFF} using a reverse engineering tool  \cite{krogmann2012ashort} and map code statements to their \ac{SEFF} element (e.g., internal actions). 
Then, they create the required probes (e.g., service probe, internal action probe, loop probe or branch probe) that refer to the \ac{SEFF} elements whose code statements have been changed. 
We define the following monitoring record types that are related to the aforementioned probe types  using \ac{IRL}  (cf. Section~\ref{kieker}).

\begin{itemize}
\item Service call record to monitor the following:\\
- the input parameters properties (e.g., type, value, number of list elements, etc.) that should be considered later as candidates for parametric dependency investigation,\\
- the caller of this service execution to learn the parametric dependency between the input parameters of both caller and callee services.\\ 
- the allocation context that captures where the component offering this service is deployed.
\item Internal-action record type to monitor the response time of the internal actions.
\item Loop record to monitor the number of loop iterations.
\item Branch record to monitor the selected branch.
\end{itemize} 
More details about the records types are  in \cite[Chapter~4.3.3]{jaegers}.

Finally, we implemented a model-based instrumentation to generate the instrumented source code as a \vitruv view. This view combines the information from two models: the source code model and the \ac{IM}.
The instrumentation starts with generating the instrumentation code for each probe in the IM according to the probe type. Then it injects the instrumentation codes into a copy of the source code. To detect the correct places for the instrumentation codes, the instrumentation process uses the relations stored in \vitruv correspondence model, i.e., the relation between probes and SEFF elements, the relation between SEFF elements and their source code statements and relation between the original source code and the generated copy (instrumented source code).

\vspace{-0.1 cm}
\section{Incremental Calibration}
\label{incremental calibration}
The following subsections explain the calibration types.
\vspace{-0.1 cm}
\subsection{\ac{Dev-time} calibration}
\label{Dev-time}
The \ac{Dev-time} changes that we consider in this paper are the source code changes that may have an impact on performance, i.e., changes in a method body.

On one hand the \textit{incremental} calibration of the SEFF branches, loops or external call arguments requires only to instrument the related source code and to analyze the resulting measurements (e.g., loop iteration number, the selected branch transition, and the values of external call parameters). The goal of this analysis is to detect whether there are dependencies to the service input parameters and express these PMPs sequentially as stochastic expression (cf. \Cref{loop,branches,arguments}).

On the other hand, the incremental calibration of the internal actions with \ac{RD} is challenging because we aim to estimate the \ac{RD} of internal actions incrementally without high monitoring overhead. The existing \ac{RDE} approaches either estimate the RDs at the service level  \cite{spinner2015evaluating} or require expensive fine-grained monitoring \cite{Brosig2009Automated,krogmann2009ck}. Therefore, we propose in the following paragraph a light-weight \ac{RDE} process that is based on adaptive instrumentation and monitoring to allow for an incremental \ac{RDE}.

Our incremental \ac{RDE} aims to estimate the \ac{RDs} in the case of adaptive monitoring, i.e., monitoring the changed parts of source code. In our running example, monitoring the code of ‘trainForRecommender’ to reestimate  its RD. For this goal, we extend the approach of Brosig et al.  \cite{Brosig2009Automated}.   
\paragraph {Basis: Non-incremental estimation of resource demands}
Brosig et al. approximate the \ac{RDs} with measured response times in the case of low resource utilization, typically 20\%. Otherwise, they estimate the \ac{RD} of internal action \(i\) (a part of SEFF, see \Cref{palladio}) for resource \(r\) (\(D_{i,r}\)) based on service demand law \cite{menasce2004} shown in equation (1). Here, \(U_{i,r}\) the average utilization of resource r due to executing internal action \(i\) and \(C_{i}\) is the total number of times that internal action \(i\) is executed during the observation period of fixed length \(T\):

\begin{equation}
D_{i,r} = \frac{U_{i,r}}
{{C_{i}}/{T}} = \frac{U_{i,r} \cdot T }
{C_{i}}
\end{equation}
Brosig et al. measure the \(C_{i}\) and estimate \(U_{i,r}\) by using the weighted response time ratios of the total resource utilization, which is not applicable in our adaptive case where not all internal actions are monitored. Therefore, we extend their approach to estimate \(U_{i,r}\) and as a result \(D_{i,r}\) based on the available measurements and the old \ac{RDs} estimations.

\paragraph{Incremental estimation of resource demands}
\label{ird}
Our new approach distinguishes internal actions into two categories based on whether they have been modified in the source code commit preceding the incremental calibration. We denote internal actions whose corresponding code regions have been modified in the preceding source code commit as \ac{MIAs}, e.g., '\textit{trainForRecommender}' in our running example, -- for these code regions, the consistency rules will generate instrumentation probes and the adaptive monitoring  produce monitoring results. We denote internal actions whose corresponding code regions have not been changed in the preceding source code commit as \ac{NMIAs}, e.g., '\textit{preprocess}' in our running example -- monitoring data for these code regions has already been observed in a previous iteration and, consequently, we have already an estimation of their RDs.

 Based on the fact that the total utilization \(U_{r}\) is measurable and the utilization due to executing \ac{NMIAs} can be estimated based on the old estimations of RDs, we can estimate \(U_{r, MIAs}\) and estimate the RD for each internal action \(i \in MIAs\) accordingly as it will be explained in following paragraphs.


To estimate (\(U_{r, NMIAs}\)),  we  estimate which internal actions \(nmi \in NMIAs\) are processed in this interval and how many times $nmi$ are called (\(C_{nmi}\)). For that, we analyze the service call records (see \Cref{adaptive instrumentation}) to determine which services are called in an observation period \(T\) and which parameters are passed. Then we traverse the service's control flows (i.e. their SEFFs) to get NMIAs and predict their RD using the input parameters. This requires evaluating branches and loops of the control flow to decide which branch transition we have to follow and how many times we have to handle the inner control flow of loops. 
Our calibration, adjusts the new or outdated branches and loops using the monitoring data (as will be described in Sections \ref{loop} and \ref{branches}) before starting this incremental RDE. Thus, we make sure that we can traverse the SEFFs control flow. Consequently, we can sum up the predicted RDs for all calls of the \ac{NMIAs} and divide the result by \(T\) to  estimate the \(U_{r,NMIAs}\) based on the utilization law as shown in the equation 2:
\begin{equation}
U_{r,NMIAs} = \frac{ \sum\limits_{nmi \in NMIAs} \: \sum\limits_{k \leq C_{nmi}} { D_{nmi_{k},r}}}{T}
\end{equation} 

 Accordingly, we estimate the utilization due to executing the \ac{MIAs} (\(U_{r,MIAs}\))  using the measured \(U_{r}\) and the estimated \(U_{r,NMIAs}\) as shown in equation (3):
\begin{equation}
U_{r,MIAs}= U_{r}-U_{r,NMIAs} 
\end{equation}

Hence, we can estimate the utilization \(U_{i,r}\) due to executing each internal action \(i  \in MIAs\) using the weighted response time ratios as shown in equation (4), where \(R_{i}\) and \(C_{i}\) are the average response time of \(i\) and its throughput. \(R_{j}\) is the average response time of the internal action \(j \in MIAs\)  and \(C_{j}\) is the number of executing it in \(T\). 
\begin{equation}
U_{i,r} =U_{r,MIAs} . \frac{R_{i} \cdot C_{i}}
{ \sum\limits_{j \in MIAs} {R_{j}} \cdot C_{j} } 
\end{equation}

Using \(U_{i,r}\) we can estimate the resource demand for  \(i\) (\(D_{i,r}\)) based on the service demand law presented in equation (1).


In the case that the host has multiple processors, our approach uses the average of the utilizations as \(U_{r}\).

Note that we assume that each internal action is dominated by a single resource. If this is not the case, we have to follow the solution proposed by Brosig et al., to measure processing times of individual execution fragments, so that the measured times of these fragments are dominated by a single resource \cite{Brosig2009Automated}.    
To differ between the \ac{CPU} demands and disk demands, we suggest detecting the disk-based services in the first activity of CIPM using specific notation or based on the used libraries. 

 \vspace{-0.1 cm}
\subsection{Ops-time calibration}
\label{Run-time calibration}
The task of the Ops-time calibration is to update the usage models as well as the deployment model according to the run-time measurements. 
To achieve that, we use the iObserve approach, which analyzes the measurements and aggregates them to detect changes concerning the deployments and the user behavior.
For this, we extended our monitoring records so that all information required by iObserve is available. This allows us to integrate the usage model extraction and the identification of deployments from iObserve. We do not need the \ac{RAC}  from iObserve, because the mapping information is implicitly presented in our monitoring records, e.g., the records that track the execution of a service include explicitly the ID of the associated service in the architectural model.
\vspace{-0.1 cm}
\section{Parametric dependencies}
\label{parametric dependencies}
This work estimates how \ac{PMPs} depend on input data and their properties (such as number of elements in a list or the size of a file). 
We begin by estimating the dependencies of branches and loops, because the incremental \ac{RDE} require traversing the SEFF control flow to estimate the utilization of \ac{NMIAs}.
Currently, we investigate a linear, quadratic, cubic and square root relations using Weka library \cite{Hall2009}. However, we are working on optimizing our estimation using genetic algorithms. The following sections   explain the analysis that we follow to estimate the parametric dependencies of \ac{PMPs}.

\vspace{-0.1 cm}
\subsection{Resource demands}
\label{rd}
To learn the parametric dependency between the resource demand of an internal action \(i\) and input parameters \(P\), we first estimate the resource demand on resource \(r\) for each combination of the input parameters (\(D_{i,r}(P)\)) using the proposed incremental RDE as described in \Cref{ird}. 


Second, we adjust the estimated \ac{RDs} of an internal action using the processing rate of the resource, where it is executed, to extract the resource demand independently of the resource' processing rate  \(D_{i}(p)\). 

Third, if the input parameters include enumerations, we perform additional analysis to test the relation between RDs and enumeration values using decision tree. If a relation is found, we build a data set for each enumeration value. Subsequently, we perform the regression analysis as it will be described in the following paragraph. Otherwise, we create one data set for each internal action that includes the estimated \ac{RDs} and their related numeric parameters. The goal is to find the potential significant relations by the regression analysis of the following equation:
\vspace{-0.15cm}
\begin{gather*}
D_{i}(P) = (a*p_0 + b*p_1 + \dotsb + z*p_n + \\
\qquad \qquad a_1*p_0^2 + b_1*p_1^2 + \dotsb + z_1*p_n^2 +\\
\qquad \qquad a_2*p_0^3 + b_2*p_1^3 + \dotsb + z_2*p_n^3  +\\ 
\qquad \qquad \qquad \qquad a_3*\sqrt{P_0} + b_3*\sqrt{p_1} + \dotsb +z_3*\sqrt{p_n}  +  \textbf{C}) \vspace{-0.1cm}
 \end{gather*}

\( p_{0}\), \( p_{1}\).. \( p_{n}\) are the numeric input parameters and the numeric attributes of objects that are input parameters.

\(a\)... \(z\), \(a_1\)... \(z_1\), \(a_2\)... \(z_2\) and \(a_3\)... \(z_3\) are the weights of the input parameters and their transformations using quadratic, cubic and square root functions. \(C\) is a constant value.

Fourth, we perform the regression analysis algorithm to find the weights of the significant relations and the constant \(C\).

Fifth, we replace the constant value \(C\) with a stochastic expression that describes the empirical distribution of \(C\) value instead of the mean value delivered by the regression analysis. This step is particularly important when no relations to the input parameters are found. In that case, the distribution function will represent the \ac{RD} of internal action better than a constant value. To achieve that, we iterate on the resulting equation that includes the significant parameters and their weights, to recalculate the value of \(C\) for each \ac{RD} value and their relevant parameters. Then, we build a distribution that represents all measured values and their frequency. 

Finally, we build the stochastic expression of \ac{RD} that may include the input parameters  and the distribution of \(C\). 

%
%
\vspace{-0.2 cm}\subsection{Loop iterations count}
\label{loop}
To estimate how the number of loop iterations depends on input parameters, we need both the loop iterations' count and the input parameters for each service call. To achieve that, we use our loop records that log the loop iterations' count, every time a loop finishes (see \Cref{adaptive instrumentation}). These records refer to the service call record that contains the input parameters.

The reason why we use additional records to count loop iterations instead of counting the total amount of enclosing service calls, is that the loop may have a nested branch or loop, which does not allow one to infer the correct count of loop iterations \cite{jaegers}.

To estimate the dependency of the loop iteration count on input parameters, we
combine the monitored loop iterations with the integer input parameter into one data
set. To do so, we filter out all non-integer parameters and take into account their integer properties like number of list elements or size of files. Then, we add transformations (quadratic and cubic) of parameters to the data set to test more relations. Finally, we use regression analysis to estimate the weights of the influencing parameters. Due to the restriction that the loop iterations count is an integer number, we have to ensure, that the output value is always an integer value, which is not always the case. Therefore, we have to approximate the non-integer weights or express them as a distribution of integer values. For instance, we can express the value 1.6 using a Palladio distribution function of an integer variable which takes the value \(1\) in \(40\%\) of all cases and the value \(2\) in \(60\%\) of cases. 
Similar to the fifth step of parametrized \ac{RDE} in \Cref{rd}, we replace the constant value of the resulting stochastic expression with an integer distribution function. This will be especially useful, when no relation to the input parameters is found. 
\vspace{-0.2 cm}
\subsection{Branch transitions}
\label{branches}
To estimate the parameterized branch transitions, we use the predefined branch monitoring records that log which branch transitions are chosen in addition to a reference to the enclosing service call.

We monitor each branch instead of predicting the selected transition according to the external call execution enclosed in the branch due to potential nested control flows (e.g., nested branches), where we cannot infer the selected transition 
\cite{jaegers}.

The used monitoring records allow us to build a data set for each branch, which includes the branch transitions and the input parameters.
To estimate the potential relations, we use the J48 decision tree of the Weka library, an implementation of the C4.5 decision tree \cite{Quinlan:1993:CPM:152181}. We filter out the non-significant parameters based on cross-validation. Finally, we transform the resulting tree into a boolean stochastic expressions for each branch transition. 
If no relation is found, the resulting stochastic expression will be a boolean distribution representing the probability of selecting a branch transition.

\vspace{-0.2 cm}\subsection{External call arguments} 
\label{arguments}
This step predicts the parameters of an external call in relation to the input parameters of the calling service.

For each parameter of an external call, we check whether it is constant, identical to one of input parameters, or depend on some of them. To identify the dependencies, we  apply linear regression in the case of numeric parameter and build a decision tree in the case of boolean/ enumeration one. For the remaining  types of parameters, we build a discrete distribution.

Because the relation between input parameters and external calls parameters may be more complex, we are working on optimizing our estimation  using a genetic search similar to the work of Krogmann et al. \cite{krogmann2009ck}.

\newcommand{\cocometime}{60 min}

\newcommand{\cocomerecords}{220706}

\newcommand{\cocomeloading}{4.664s}

\newcommand{\cocomecalibration}{9.168s}

\newcommand{\cocomeusage}{2.365s}

\newcommand{\cocomeselfvalidation}{1.729s}

\newcommand{\cocomeoverhead}{\textbf{1.31ms} }
\newcommand{\cocomeoverheadsec}{\textbf{252\boldmath$\mu$s} }

\newcommand{\cocomemin}{0.1248}
\newcommand{\cocomemax}{0.1774}
\newcommand{\cocomeavg}{0.1467}
\newcommand{\teastoretime}{60 min}

\newcommand{\teastorerecords}{294237}

\newcommand{\teastoreloading}{5.142s}

\newcommand{\teastorecalibration}{4.996s}

\newcommand{\teastoreusage}{0.194s}

\newcommand{\teastoreoverhead}{\textbf{1.81ms} }
\newcommand{\teastoreoverheadsec}{\textbf{88\boldmath$\mu$s} }

\newcommand{\teastoremin}{0.1298}
\newcommand{\teastoremax}{0.3875}
\newcommand{\teastoreavg}{0.2308}

\newcommand{\teastoreselfvalidation}{1.440s}

\newcommand{\teastoreitZrecords}{27085}

\newcommand{\teastoreitZloading}{0.774s}

\newcommand{\teastoreitZcalibration}{0.789s}

\newcommand{\teastoreitZusage}{0.165s}

\newcommand{\teastoreitZselfvalidation}{1.327s}
\vspace{-0.1 cm}
\begin{figure*}[!h]
	\centering
    \includegraphics[width=0.8\textwidth]{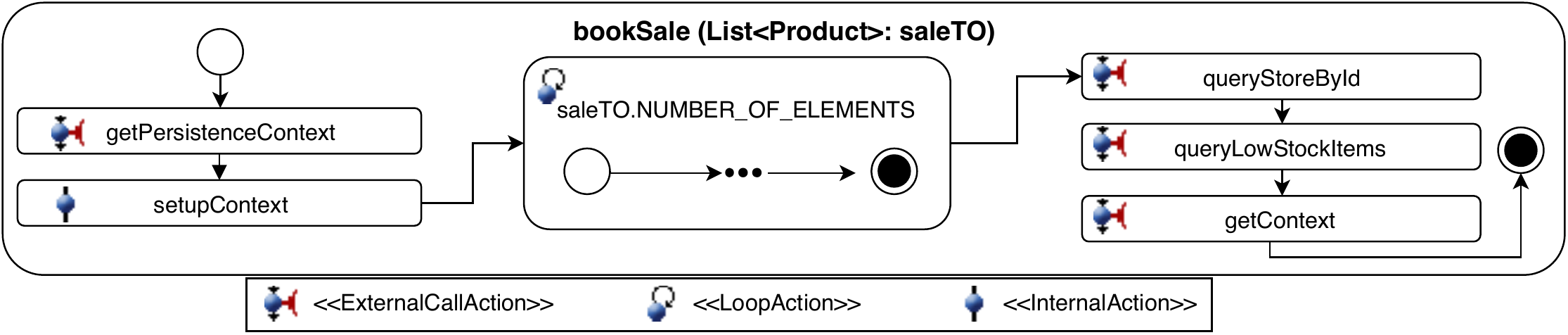} 
    \vspace{-0.2 cm}
    \caption{\ac{SEFF} of CoCoME's "bookSale" service}
    \label{fig:seff:booksale}
\end{figure*}

\section{Evaluation}
\label{evaluation}
This section introduces the evaluation goals, setup, environment and the results of the evaluation:
\vspace{-0.3 cm}
\subsection{Evaluation Goals}
\label{goals}
The goal is to evaluate the following three aspects of the approach and answer the  Evaluation-Questions (EQs):
\begin{itemize}[leftmargin=*]
\item \textbf{Accuracy of the incremental calibration}: \\
\textbf{EQ1.1:} How accurate are the incremental calibrated PMs?
\\
\textbf{EQ1.2:} Is the accuracy of the incremental calibrated PM stable over the incremental evolution? 
\item \textbf{Accuracy of the parametric dependencies}: \\
\textbf{EQ2:} Does the estimation of parametric dependencies improve the accuracy of PMs?
\item \textbf{Performance of the calibration pipeline:} \\ \textbf{EQ3:} How long does the pipeline need to update PMs?
  \item \textbf{Monitoring overhead:} \\\textbf{EQ4.1:} How much is the required monitoring overhead?
  \\\textbf{EQ4.2:} How much can the adaptive instrumentation reduce the monitoring overhead?

\end{itemize} 
\vspace{-0.2 cm}
\subsection{Experiment Setup}
\label{evaluation:setup}
The evaluation is structured as follows:\\
    1. We monitor the application over a defined period of
    time (e.g., 60 minutes) and perform a load test at the same time to artificially simulate user interactions. \\
    2. We run the monitoring in combination with the load test two times. First, we apply a fine-grained monitoring, pass the results to the
    calibration pipeline for adjusting the architecture model (training set). Second, we only observe service calls (coarse-grained monitoring). These data are used to estimate the accuracy of the calibrated model (validation set). This procedure prevents the monitoring overhead from falsifying the evaluation results. We also record the execution times (overhead) of the pipeline parts.
    3. We adjust the usage model according to the validation set and perform a fixed number of simulations of the updated architecture
    model (100 repetitions). Repetitions are necessary, as the simulation samples from stochastic distributions may not be representative. Afterward, we compare the results of the
    simulations with the validation set. Both are distributions,
    therefore we use the \ac{KS-Test} \cite{KSTest}, the Wasserstein metric \cite{majewski_et_al:LIPIcs:2018:9327}, and conventional statistical measures to determine the 
    similarity. 
    The lower the \ac{KS-Test} value, the higher is the accuracy of the updated architecture model. In order to limit the dependence on a single metric, we also included the Wasserstein distance to confirm the results. The Wasserstein metric is distance measure that quantifies the effort needed to transfer one distribution into the other. If we want to compare the accuracy of two simulation distributions (using real monitoring as reference), this metric is well suited, but reports absolute numbers that  are  difficult to interpret. We therefore combined both metrics in our analysis.
\vspace{-0.3 cm}
\subsection{Evaluation Environments}
We evaluated our approach using two case studies: \ac{CoCoME}~\cite{cocome} and
TeaStore \cite{teastore}. 
 \ac{CoCoME}  is a trading system which is designed for the use in supermarkets. It supports several processes like scanning products at a cash desk or processing sales using a credit card. We used a cloud-based implementation of \ac{CoCoME} where the enterprise server and
the database run in the cloud.

Additionally, we evaluated our approach using the aforementioned running example of the TeaStore case study (cf. \Cref{Running Example}). Compared to the original, we modified the application so that the "train" service is executed when a user places an order. This causes the number of users to have a direct impact on the service response times. 

Because \ac{CoCoME} and TeaStore are not implemented using the \vitruv platform, we instrument the changed parts of the source code manually. In future work, we will use approaches for importing existing source code into \vitruv \cite{leonhardt2015a}, which allows the automatic adaptive instrumentation.

\vspace{-0.2cm}
\subsection{Evaluation of the Incremental Calibration }
\label{Evaluation incremental Calibration}
This section evaluates the accuracy of our calibration for the following incremental evolution scenarios.
\subsubsection{BookSale Calibration}
The ``bookSale'' service of CoCoMe case study is responsible for processing a sale after a user submitted his payment. As input, this service receives a list of the purchased products including all information related to the purchase. The service consists of several internal and external actions and two loops.
Figure \ref{fig:seff:booksale} visualizes the structure of the calibrated "bookSale" service
of \ac{CoCoME}. For reasons of simplification, we focused on the structure and omitted certain details. 
In this evaluation scenario, we supposed that the ``bookSale'' service  is newly added.
According to our assumption, we instrumented this service and all services that are subsequently called by it. 
We performed a load test which simulates user purchases. We used the resulting monitoring data to calibrate our architecture model.

\paragraph{Results}  Table \ref{table:cdf} shows the quartiles for both the monitoring and the simulation results for a single example execution. Besides, a KS test value of \textbf{0.1321} and a Wasserstein distance of \textbf{13.7726} were obtained by comparing the monitoring and the simulation distributions. It is well visible that these  distributions are very close to each for the first, second and third quartile. The discrepancy in the minimum and maximum values of the distributions arises because we limited the time spent for simulation.  The KS tests average over all 100 simulation runs is approximately \textbf{\cocomeavg} (min \textbf{\cocomemin}, max \textbf{\cocomemax}). These results confirm the PM accuracy (EQ1.1).
\vspace{-0.1cm}

\renewcommand{\arraystretch}{1.4}
\begin{table}[!h]
    \centering
    \caption{Quartiles for the probability distributions of a single simulation and the monitoring for the "bookSale" service}
    \begin{tabularx}{\columnwidth}{p{1.6cm} X X X X X X}
    \hline 
    Distribution & Min & Q1 & Q2 & Q3 & Max & Avg \\ \hline
    Monitoring & 31ms & 101ms & 130ms & 187ms & 461ms & 143ms \\
    Simulation & 56ms & 95ms & 124ms & 163ms & 338ms & 133ms \\ \hline
    \end{tabularx}
    \label{table:cdf}
\end{table}

\renewcommand{\arraystretch}{1.0}
\vspace{-0.1 cm}

\subsubsection{Train Calibration}
\label{Train Calibration}
Similar to ”BookSale” evaluation scenario, we assumed that ’train’ service (cf. \Cref{Running Example})  was newly added. Therefore, we instrumented and monitored the statements of all SEFF elements shown in Fig. \ref{fig:seff:train} in two cases: (1) for the first implementation of ’train’ and (2) after adding an enumeration parameter that refers to the used implementation of 'train' to represent the four implementations in one SEFF (cf. Sec. \ref{whatIf}). Then we calibrated the  SEFFs of these two cases. Due to lack of space we show only the accuracy results of the second case in (Table \ref{table:train:kss}, \nth{2} column)\footnote{See \url{https://sdqweb.ipd.kit.edu/wiki/CIPM} for more detail on evaluation.}.
\subsubsection{TrainForRecommender Calibration}
\label{TrainForRecommender}
To evaluate that the accuracy of PM is stable (EQ1.2), we simulated an evolution scenario that assumes that the developers change the implementation of the \textit{'train'} service to examine the four implementation alternatives (cf. \Cref{Running Example}) in four subsequent source code commits $c1$, ..., $c4$. According to our assumption, except of the "trainForRecommender" internal action, the behavior  of \textit{'train'} (\Cref{fig:seff:train}) stays the same.
To evaluate this scenario, we instrumented only the "trainForRecommender" internal action for the first implementation of \textit{'train'} ($c1$). 
We performed 15 minutes of fine-granular monitoring and used the resulting data to calibrate the model. Next, we monitored the 'train' service for another 15 minutes of coarse granular. The coarse granular monitoring data was used as reference for the estimation of the accuracy of PM. In the next step,  we instrumented the "trainForRecommender" internal action for the second implementation of \textit{'train'} ($c2$) and performed a fine granular and a coarse granular monitoring for 15 minutes each. We repeated the same procedure for commits $c3$ and $c4$. 

To simulate additional commits, we changed the train implementation by randomly replacing the strategy with another one. We repeated this procedure six more times to simulate six additional iterations and calculate the accuracy of the calibrated model for each iteration. In addition, the database filled up over time, which leads to increasing response times, increasing the difficulty for the calibration process.
\vspace{-0.15 cm}
\paragraph{Results}

\vspace{-0.1 cm}
\begin{table}[!h]
\caption{Aggregated metrics of the evolution scenario}
    \begin{tabularx}{\columnwidth}{p{1.6cm} |X X X | X  X}
    \hline 
    Metric & Min & Median & Max & Avg  & St.Dev \\ \hline
    KS Test & 0.222  & 0.257 & 0.317  & 0.256 &  0.028   \\
    WS Distance & 24.56 &  33.74 & 38.96 &  33.76& 4.01  \\ \hline
    \end{tabularx}
     \label{table:evol}
\end{table}
\vspace{-0.15 cm}
Table \ref{table:evol} lists the aggregated results. 
The values are slightly higher than without applying changes to the application. However, it can be seen that the accuracy of the models is almost constant throughout the evolutionary steps. From this, we conclude that the automated calibration is able to handle changes of the observed services.
\vspace{-0.25cm}
\subsection{Evaluation of the Estimation of Parametric Dependencies}
 \label{whatIf}
This scenario  evaluates the recognition of the parametric dependencies and compares the accuracy of the AbPP using the parameterized model (i.e., calibrated with our approach) with the non-parameterized model (i.e., calibrated ignoring parametric dependencies). For this goal, we extended the SEFF shown in \Cref{fig:seff:train} and added an enumeration input parameter that refers to the recommender type and determines which implementation is used within the "trainForRecommender" internal action. This allowed us to represent all different implementations in one SEFF. This means the RD of "trainForRecommender" can be related to the recommender type. Then, we calibrated the SEFF and predicted the performance for increasing the load from 20  to 40 concurrent users.  
First, we monitored the service fine granular (F1) and subsequently coarse granular (C1) by applying the load of 20 users. We \textbf{only} used the resulting  monitoring data (F1) to calibrate a parameterized model using our approach and the non-parameterized model using the distributions of the observed values. Then, we compared the simulation results of both models with the monitoring data (C1). Then, we performed AbPP for the higher load (40 users). As baseline, we measured the performance using coarse granular monitoring and applying the load of 40 users (C2).  Finally, we compared the AbPP results  of both models with the monitoring data (C2).
\paragraph{Results}
The results confirmed that our calibration was able to detect the parametric dependencies. Hence, the RD of \textit{trainForRecommender} is related to the type of recommendation implementation and the number of elements.

Table~\ref{table:train:kss} shows the comparison of the KS tests and the Wasserstein distances between the parameterized and the non-parameterized model. It can be seen that both models are very accurate for the standard load of 20 users. However, the non-parametric model loses significantly more accuracy than the parameterized model, when the load changes to 40 users. This confirms that the estimation of the parametric dependencies improves the accuracy of PMs (EQ1.1, EQ2).
\vspace{-0.15 cm}
\renewcommand{\arraystretch}{1.4}

\begin{table}[!h]
    \centering
    \caption{Comparison between the accuracy of the parameterized (PM) and the non-parameterized (NPM) model.
    }
    \begin{tabularx}{\columnwidth}{p{1.2cm}|X X|X X}
    \toprule[1.0pt]
    \thead{Metric} & \thead{\shortstack{PM\\(20 Users)}} & \thead{\shortstack{NPM\\(20 Users)}} & \thead{\shortstack{PM\\(40 Users)}} & \thead{\shortstack{NPM\\(40 Users)}} \\ \midrule[1.0pt]
    KS Q1 & 0.1024 & \textbf{0.1023} & \textbf{0.1378} & 0.2352 \\
    KS Avg. & 0.1267 & \textbf{0.1239} & \textbf{0.1609} & 0.2575 \\
    KS Q3 & \textbf{0.1431} & 0.1441 & \textbf{0.1810} & 0.2834 \\ \hline
    WS Q1 & 15.4911 & \textbf{11.2484} & \textbf{33.0235} & 43.5959 \\
    WS Avg. & 19.1764 & \textbf{16.2669} & \textbf{39.6138} & 51.3531 \\
    WS Q3 & 22.2654 & \textbf{18.2179} & \textbf{46.8370} & 59.9197 \\
    \hline
    \end{tabularx}
    \label{table:train:kss}
\end{table}
\vspace{-0.1 cm}
\renewcommand{\arraystretch}{1.0}

\renewcommand{\arraystretch}{1.0}	

\vspace{-0.2 cm}
\subsection{Evaluation of the  MbDevOps Pipeline}
This section evaluates the performance of the implemented part of MbDevOps Pipeline (monitoring, incremental calibration and self-validation) and answers EQ3. 
For that, we computed how long the individual pipeline parts took to complete. We used the three evolution scenarios described in \Cref{Evaluation incremental Calibration}. The results are shown in Table \ref{table:exec}.
\vspace{-0.15 cm}\begin{table}[h]
    \centering
    \caption{Execution times of the pipeline parts}
    \begin{tabular}{|c|c|c|c|}
    \hline 
    Measure & bookSale & train  & trainForRecommender  \\ \hline \hline
    Record Count & \cocomerecords & \teastorerecords & \teastoreitZrecords \\ \hline \hline
    Load Records & \cocomeloading & \teastoreloading & \teastoreitZloading \\ \hline
    PM Calibration & \cocomecalibration & \teastorecalibration & \teastoreitZcalibration \\ \hline
    UM Adjustment & \cocomeusage & \teastoreusage & \teastoreitZusage \\ \hline
    Self-Validation & \cocomeselfvalidation & \teastoreselfvalidation & \teastoreitZselfvalidation \\ \hline \hline
    \textbf{Total} & 17.926s & 11.772s & 3.055s \\ \hline
    \end{tabular}
    \label{table:exec}
\end{table}
\vspace{-0.15 cm}

We can see that the required time strongly depends on the number of monitoring records and the complexity of the observed service. Except for the self-validation, the execution time of all parts of the pipeline depends on the number of records (see Table \ref{table:perf}). In all considered cases, the execution of the whole pipeline took less than \textbf{20} seconds. 

We observe that adaptive monitoring limits the number of monitoring records. Accordingly, the second iteration of TeaStore monitoring produces far fewer records, which results in significantly lower execution times.
To gain detailed insights about the performance of the pipeline, we also examined the behavior for an increasing number of monitoring records. For this purpose, we generated monitoring data using \ac{CoCoME}  and the load test that we also used before. Thereafter, we executed the pipeline  several times, with an increasing number of monitoring records as input. The results in Table \ref{table:perf} showed that the execution time of the pipeline scales linearly with the number of monitoring records in this case.
\renewcommand{\arraystretch}{1.2}
\newcolumntype{Y}{>{\centering\arraybackslash}X}
 \vspace{-0.1 cm}
\begin{table}[!h]
    \centering
    \caption{Detailed performance information for an increasing number of monitoring records}
    \begin{tabularx}{\columnwidth}{|X|X X X|Y|}
    \hline
    Number of records & Loading Records & PM Calibration & UM Adjustment & Sum  \\ \hline
    100000 & 1.570s & 6.899s & 1.158s & \textbf{9.627s} \\
    200000 & 3.018s & 7.229s & 1.860s & \textbf{12.107s} \\
    500000 & 6.290s & 7.899s & 1.802s & \textbf{15.991s} \\
    700000 & 9.972s & 8.194s & 1.942s & \textbf{20.108s} \\
    1000000 & 14.780s & 9.160s & 1.917s & \textbf{25.857s} \\ \hline
    \end{tabularx}
    \label{table:perf}
   
\end{table}
 \vspace{-0.1 cm}
\renewcommand{\arraystretch}{1.0}
 \vspace{-0.1 cm}
\vspace{-0.25 cm}
\subsection{Evaluation of Monitoring Overhead}
To answer the EQ4.1 and EQ4.2, we measured the overhead caused by the monitoring for the incremental calibration scenarios in \Cref{Evaluation incremental Calibration}. 

The average monitoring overhead for \textit{bookSale}  was \cocomeoverhead (fine-grained), \cocomeoverheadsec (coarse-grained) and for \textit{train}  \teastoreoverhead (fine-grained), \teastoreoverheadsec (coarse-grained). We note that the coarse-grained monitoring has a negligible influence. The fine-grained monitoring overhead scales with the complexity of the service. In our case, the overhead of around 1ms had no drastic impact on the performance, since the execution times of the services are significantly higher.

The evolution of \textit{TrainForRecommender} calibration scenario (cf. \Cref{TrainForRecommender}) shows very well that the adaptive monitoring helps to significantly reduce the monitoring overhead.
Comparing to \textit{train} (cf. \Cref{Train Calibration}) where all SEFF elements are instrumented, we saved \textbf{at least 75\%} of the monitoring overhead by instrumenting only the changed internal action.
This is reflected in the calibration times, which are considerably lower compared to the first iteration (see \autoref{table:exec}) (EQ4.2).
In general, there is a trade off between the monitoring overhead and the accuracy of the model. However, our approach tries to balance them by using the incremental calibration in combination with the adaptive monitoring in order to minimize calibration time and monitoring overhead.

\section{Related Work} 
\label{related works}
A number of approaches for constructing the architectural model based on static (e.g., \cite{becker2010reverse}), dynamic, or hybrid analysis exist.
Walter et al. \cite{WaStKoKo2017-QUDOS-PMXBuilder} propose a  tool to extract an architectural PM as well as performance annotation based on analysing monitoring traces.
Similarly, other works \cite{Brunnert13,spinner2016reference,BrHuKo2011-ASE-AutomExtraction} extract PM based on dynamic analysis. Krogmann et al.  \cite{krogmann2009ck,krogmann2012ashort} extract parametrized PCM  based on hybrid analysis.
Langhammer et al. introduce two reverse engineering tools \cite [P.~140]{langhammer2017a} \cite{langhammer2016automated}  that extract the behavior  of the underlying source code based on hybrid analysis.

The above-mentioned approaches do not support the iterative extraction of  the architectural PMs or the incremental calibration. If they were used in an iterative development, they would re-extract and recalibrate the whole  models after each iteration. This would cause a high monitoring overhead. Moreover, this process ignores the potential manual modifications of the extracted PM that may be formerly applied.


 In addition to the work of Krogmann et al. \cite{krogmann2009ck,krogmann2012ashort}, the works of Ackermann et al.~\cite{AcGrEiKo2018-DependencyModeling} and Curtois et al.~\cite{Courtois2000}  also consider the characterization of parametric dependencies in performance models, while Grohmann et al.~\cite{GrEiElMaKiKo2019-MASCOTS-DependencyIdentification} focus on the identification of those from monitoring data. However, our work considers the parametric dependencies during the incremental calibration of the architectural PMs.
 
Model extraction approaches derive the resource demands either based on coarse-grained monitoring data \cite{spinner2015evaluating,LibReDE} or fine-grained data 
\cite{Brosig2009Automated,willnecker2015comparing}. The latter approaches give a higher accuracy by estimating \ac{PMPs} but have a downside effect because of the overhead of instrumentation and the monitoring. Our approach reduces the overhead by the automatic adaptive instrumentation and monitoring.

Spinner et al. \cite{SpGrEiKo2019-JSS-ModelLearning} propose an agent-based approach to updated architectural performance models. In contrast to our work, they focus on model updates at run-time.

\ac{DPE} \cite{WaEiGrOkKo2018-ICPE-Tools-for-DPE-Tutorial} technically enables to answer concerns based on measurement-based performance evaluation and model-based performance predictions. However, existing solutions do not provide a technical integration into the build pipeline. Also, \ac{DPE} does not answer how to keep PMs updated.

\vspace{-0.1cm}
\section{Conclusion and future works}
\label{Conclusion}
Applying \ac{AbPP} in the agile and DevOps process promises to detect performance problems by simulation instead of the execution in a production environment. 
We presented the continuous integration of the architectural performance model in the development process based on the static and dynamic analysis of the changed code. Our approach keeps the PM continuously up-to-date. We also presented the \ac{MbDevOps} pipeline which automates the incremental calibration process. 

Our calibration estimates the parametrized PMPs incrementally and uses a novel incremental resource demand estimation based on adaptive monitoring. Moreover, our calibration updates both the usage and deployment models to support model-based run-time performance management like auto-scaling.

We tested the functionality using two case studies. 
The evaluation showed the ability to calibrate the \ac{PM} incrementally. We evaluated the accuracy of our calibration by comparing the simulation results with the monitoring data. In these case studies, the incrementally calibrated models and their parametric dependencies were accurate and the used adaptive monitoring significantly reduced the monitoring overhead and the calibration effort.

In this work, we have automated only the part of the proposed \ac{MbDevOps} pipeline that is responsible for testing, incremental calibration and self-validation.

In future work, we aim to automate the whole pipeline to integrate the \ac{CI} process with \vitruv.
 Moreover, we will extend the available approaches to integrate existing source code and models into \vitruv platform \cite{leonhardt2015a, Mazkatli2018MiSE}.

 Besides, we will extend our parametrized calibration to test for more complex dependencies of external call parameters. It is also planned to extend the monitoring records to update the system model (composition of the components) automatically. 

\printbibliography

@String{Computer = "{IEEE} Computer" }

@String{Computing = "Computing" }

@String{Springer = "Springer-Verlag" }

@InProceedings{Brosig2009Automated,
  author    = {Fabian Brosig et al.},
  title     = {{Automated Extraction of Palladio Component Models from Running Enterprise Java Applications}},
  booktitle = {Proceedings of the 1st Int. Workshop on Run-time models for Self-managing systems and applications. In conjunction with the Fourth Int. Conference on Performance Evaluation Methodologies and Tools (VALUETOOLS 2009)},
  year      = {2009},
  }

@InProceedings{kieker,
  author    = {van Hoorn, Andr{\'e} and Waller, Jan and Hasselbring, Wilhelm},
  title     = {Kieker: A Framework for Application Performance Monitoring and Dynamic Software Analysis},
  booktitle = {Proceedings of the 3rd ACM/SPEC Int. Conference on Performance Engineering},
  year      = {2012},
  series    = {ICPE '12},
  publisher = {ACM},
  pages     = {247--248},
  acmid     = {2188326},
   numpages  = {2},
}

@Book{krogmann2012ashort,
  author    = {Klaus Krogmann},
  title     = {{Reconstruction of Software Component Architectures and Behaviour Models using Static and Dynamic Analysis}},
  year      = {2012},
  volume        = {4},
  series    = {The Karlsruhe Series on Software Design and Quality},
  publisher = {{KIT Scientific Publishing}},
  series-editor = {Ralf Reussner},
}

@Book{menasce2004,
  author    = {Menasce, Daniel A and Almeida, Virgilio AF and Dowdy, Lawrence W and Dowdy, Larry},
  title     = {Performance by design: computer capacity planning by example},
  year      = {2004},
  publisher = {Prentice Hall Professional},
}

@Article{spinner2015evaluating,
  author    = {Spinner, Simon and Casale, Giuliano and Brosig, Fabian and Kounev, Samuel},
  title     = {Evaluating approaches to resource demand estimation},
  journal   = {Performance Evaluation},
  year      = {2015},
  volume    = {92},
  publisher = {Elsevier},
}

@article{SpGrEiKo2019-JSS-ModelLearning,
  title = {Online model learning for self-aware computing infrastructures},
  journal = {Journal of Systems and Software},
  volume = {147},
  pages = {1 - 16},
  year = {2019},
  issn = {0164-1212},
    author = {Simon Spinner and Johannes Grohmann and Simon Eismann and Samuel Kounev},
}

@InProceedings{willnecker2015comparing,
  author       = {Willnecker, Felix and Dlugi, Markus and Brunnert, Andreas and Spinner, Simon and Kounev, Samuel and Gottesheim, Wolfgang and Krcmar, Helmut},
  title        = {Comparing the accuracy of resource demand measurement and estimation techniques},
  booktitle    = {European Workshop on Performance Engineering},
  year         = {2015},
  organization = {Springer},
}

@Article{aop,
  author    = {Kiczales, Gregor and Lamping, John and Mendhekar, Anurag and Maeda, Chris and Lopes, Cristina and Loingtier, Jean-Marc and Irwin, John},
  title     = {Aspect-oriented programming},
  journal   = {ECOOP 97—Object-oriented programming},
  year      = {1997},
  pages     = {220--242},
  publisher = {Springer},
}

@InProceedings{LibReDE,
  author    = {Spinner, Simon and Casale, Giuliano and Zhu, Xiaoyun and Kounev, Samuel},
  title     = {LibReDE: A Library for Resource Demand Estimation},
  booktitle = {Proceedings of the 5th ACM/SPEC Int. Conference on Performance Engineering},
  year      = {2014},
  series    = {ICPE '14},
  publisher = {ACM},
  }

@Book{conni,
  author    = {Connie U. Smith and Lloyd G. Williams},
  title     = {Performance Solutions: A Practical Guide to Creating Responsive, Scalable Software},
  year      = {2003},
  publisher = {Addison Wesley Longman Publishing Co., Inc.},
  isbn      = {0-201-72229-1},
  address   = {Redwood City, CA, USA},
}

@InProceedings{spinner2016reference,
  author    = {Spinner, Simon and Walter, J\"{u}rgen and Kounev, Samuel},
  title     = {A Reference Architecture for Online Performance Model Extraction in Virtualized Environments},
  booktitle = {Companion Publication for ACM/SPEC on Int. Conference on Performance Engineering},
  year      = {2016},
  organization = {ACM},
  pages     = {57--62},
  
}

@Article{krogmann2009ck,
  author    = {Klaus Krogmann and Michael Kuperberg and Ralf Reussner},
  title     = {{Using Genetic Search for Reverse Engineering of Parametric Behaviour Models for Performance Prediction}},
  journal   = {IEEE Transactions on Software Engineering},
  year      = {2010},
  publisher = {{IEEE}},
}

@TechReport{brunnert2015a,
  author      = {Andreas Brunnert and Andr{\'{e}} van Hoorn and Felix Willnecker and Alexandru Danciu and Wilhelm Hasselbring and Christoph Heger and Nikolas Roman Herbst and Pooyan Jamshidi and Reiner Jung and J{\'{o}}akim von Kistowski and Anne Koziolek and Johannes Kro{\ss} and Simon Spinner and Christian V{\"{o}}gele and J{\"{u}}rgen Walter and Alexander Wert},
  title       = {Performance-oriented DevOps: {A} Research Agenda},
  institution = {SPEC Research Group - DevOps Performance Working Group, Standard Performance Evaluation Corporation (SPEC)},
  year        = {2015},
  number      = {SPEC-RG-2015-01},
  month       = aug,
  url         = {http://arxiv.org/abs/1508.04752},
  bibsource   = {dblp computer science bibliography, http://dblp.org},
}

@InProceedings{Mazkatli2018Qudos,
  author    = {Mazkatli, Manar and Koziolek, Anne},
  title     = {Continuous Integration of Performance Model},
  booktitle = {Companion of the 2018 ACM/SPEC Int. Conference on Performance Engineering},
  year      = {2018},
  series    = {ICPE '18},
  publisher = {ACM},
  location  = {Berlin, Germany},
  isbn      = {978-1-4503-5629-9},
  
  doi       = {10.1145/3185768.3186285},
  url       = {http://doi.acm.org/10.1145/3185768.3186285},
  acmid     = {3186285},
  address   = {New York, NY, USA},
  numpages  = {6},
  tags      = {Vitruv},
}

@Thesis{jaegers,
  author      = {Jan-Philipp Jägers},
  title       = {Iterative Performance Model Parameter Estimation Considering Parametric Dependencies},
  type        = {Master Thesis},
  institution = {Karlsruher Institut für Technologie Fakultät für Informatik},
  year        = {2018},
  date        = {2019-05-13},
}

@Article{mbpp,
  author  = {S. {Balsamo} and A. {Di Marco} and P. {Inverardi} and M. {Simeoni}},
  title   = {Model-based performance prediction in software development: a survey},
  journal = {IEEE Transactions on Software Engineering},
  year    = {2004},
  volume  = {30},
  number  = {5},
  month   = {5},
  pages   = {295-310},
  issn    = {0098-5589},
  doi     = {10.1109//TSE.2004.9},
}

@InProceedings{SPE,
  author    = {Pooley, Rob},
  title     = {Software engineering and performance: a roadmap},
  booktitle = {ICSE-Future of SE Track},
  year      = {2000},
  pages     = {189--199},
}

@Article{SPEwoodside,
  author  = {C. Murray Woodside et al.},
  title   = {The Future of Software Performance Engineering},
  journal = {Future of Software Engineering (FOSE '07)},
  year    = {2007},
  pages   = {171-187},
}

@Article{cocome,
  author  = {R. Heinrich and S. G{\"a}rtner  and T. Hesse and T. Ruhroth and R. Reussner and K. Schneider and B. Paech and J. J{\"u}rjens},
  title   = {The {CoCoME} Platform: A Research Note on Empirical Studies in Information System Evolution},
  journal = {Int. Journal of Software Engineering and Knowledge Engineering},
  year    = {2015},
  volume  = {25},
  number  = {09\&10},
  pages   = {1715-1720},
  doi     = {10.1142/S0218194015710059},
  
}

@InProceedings{teastore,
  author       = {von Kistowski, J{\'o}akim and Eismann, Simon and Schmitt, Norbert and Bauer, Andr{\'e} and Grohmann, Johannes and Kounev, Samuel},
  title        = {TeaStore: A Micro-Service Reference Application for Benchmarking, Modeling and Resource Management Research},
  booktitle    = {2018 IEEE 26th Int. Symposium on Modeling, Analysis, and Simulation of Computer and Telecommunication Systems (MASCOTS)},
  year         = {2018},
  organization = {IEEE},
  pages        = {223--236},
}

@Article{ci1,
  author    = {Meyer, Mathias},
  title     = {Continuous integration and its tools},
  journal   = {IEEE software},
  year      = {2014},
  volume    = {31},
  number    = {3},
  pages     = {14--16},
  publisher = {IEEE},
}

@InProceedings{cd,
  author       = {Shahin, Mojtaba},
  title        = {Architecting for devops and continuous deployment},
  booktitle    = {Proceedings of the ASWEC 2015 24th Australasian Software Engineering Conference},
  year         = {2015},
  organization = {ACM},
}

@Article{Hall2009,
  author    = {M. Hall and E. Frank and G. Holmes and B. Pfahringer and P. Reutemann and Ian H. Witten},
  title     = {The {WEKA} data mining software},
  journal   = {{ACM} {SIGKDD} Explorations Newsletter},
  year      = {2009},
  volume    = {11},
  number    = {1},
  month     = {nov},
  pages     = {10},
 
  publisher = {Association for Computing Machinery ({ACM})},
}

@Book{Quinlan:1993:CPM:152181,
  author    = {Quinlan, J. Ross},
  title     = {C4.5: Programs for Machine Learning},
  year      = {1993},
  publisher = {Morgan Kaufmann Publishers Inc.},
  isbn      = {1-55860-238-0},
  address   = {San Francisco, CA, USA},
}

@Inbook{KSTest,
author={Dodge, Yadolah},
  title="Kolmogorov--Smirnov Test",
  bookTitle="The Concise Encyclopedia of Statistics",
  year="2008",
  publisher="Springer New York",
  address="New York, NY",
  pages="283--287",
  isbn="978-0-387-32833-1",
  doi="10.1007/978-0-387-32833-1_214",
  
}

@inproceedings{Courtois2000,
 author = {Courtois, Marc and Woodside, Murray},
 title = {Using Regression Splines for Software Performance Analysis},
 booktitle = {Proceedings of the 2nd Int. Workshop on Software and Performance},
 year = {2000},

 numpages = {10},
}

@inproceedings{AcGrEiKo2018-DependencyModeling,
  title = {Black-box Learning of Parametric Dependencies for Performance Models},
  author = {Ackermann, Vanessa and Grohmann, Johannes and Eismann, Simon and Kounev, Samuel},
  booktitle = {Proceedings of 13th Workshop on Models@run.time (MRT), co-located with MODELS 2018},
  year = {2018},
  series = {CEUR Workshop Proceedings},
  location = {Copenhagen, Denmark},
  month = {October},
  eventdate = {2018-10-14}
}

@inproceedings{WaEiGrOkKo2018-ICPE-Tools-for-DPE-Tutorial,
  author = {J{\"u}rgen Walter and Simon Eismann and Johannes Grohmann and Dusan Okanovic and Samuel Kounev},
  title = {{Tools for Declarative Performance Engineering}},
  titleaddon = {{(Tutorial Paper)}},
  booktitle = {Companion of the 2018 ACM/SPEC Int. Conference on Performance Engineering},
  series = {ICPE '18},
  year = {2018},
  isbn = {978-1-4503-5629-9},
  location = {Berlin, Germany},
  pages = {53--56},
  numpages = {4},
  doi = {10.1145/3185768.3185777},
  acmid = {3185777},
  publisher = {ACM},
  address = {New York, NY, USA}
}

@incollection{koziolek_modeling_2016,
	address = {Cambridge, Massachusetts},
	title = {Modeling {Quality}},
	isbn = {978-0-262-03476-0},
	booktitle = {Modeling and simulating software architectures: the {Palladio} approach},
	publisher = {MIT Press},
	author = {Koziolek, Heiko},
	year = {2016}}

@InProceedings{leonhardt2015a,
  author    = {Leonhardt, Sven and Hettwer, Benjamin and Hoor, Johannes and Langhammer, Michael},
  title     = {Integration of Existing Software Artifacts into a View- and Change-Driven Development Approach},
  booktitle = {Proceedings of the 2015 Joint MORSE/VAO Workshop on Model-Driven Robot Software Engineering and View-based Software-Engineering},
  year      = {2015},
  series    = {MORSE/VAO '15},
  pages     = {17--24},
  publisher = {ACM},
  acmid     = {2802061},
  numpages  = {8}
  }

@inproceedings{GrEiElMaKiKo2019-MASCOTS-DependencyIdentification,
  author = {Johannes Grohmann and Simon Eismann and Sven Elflein and Manar Mazkatli and J{\'o}akim von Kistowski and Samuel Kounev},
  title = {{Detecting Parametric Dependencies for Performance Models Using Feature Selection Techniques}},
  booktitle = {Proceedings of the 27th IEEE Int. Symposium on the Modelling, Analysis, and Simulation of Computer and Telecommunication Systems},
  series = {MASCOTS '19},
  year = {2019},
  month = {October},
  location = {Rennes, France}
}

@inproceedings{BeEiFeGrRhJaShHoViWaWi2019-ICPE-DevOpsSurvey,
  author = {Cor{-}Paul Bezemer and Simon Eismann and Vincenzo Ferme and Johannes Grohmann and Robert Heinrich and Pooyan Jamshidi and Weiyi Shang and Andr{{\'e}} van Hoorn and M{{\'o}}nica Villavicencio and J{{\"u}}rgen Walter and Felix Willnecker},
  title = {{How is Performance Addressed in DevOps?}},
  booktitle = {Proceedings of the 2019 ACM/SPEC Int. Conference on Performance Engineering},
  pages = {45--50},
  year = {2019}
}

@InProceedings{majewski_et_al:LIPIcs:2018:9327,
  author =	{Szymon Majewski and Michal Aleksander Ciach and Michal Startek and Wanda Niemyska and Blazej Miasojedow and Anna Gambin},
  title =	{{The Wasserstein Distance as a Dissimilarity Measure for Mass Spectra with Application to Spectral Deconvolution}},
  booktitle =	{18th Int. Workshop on Algorithms in  Bioinformatics (WABI 2018)},
  pages =	{25:1--25:21},
  series =	{Leibniz Int. Proceedings in Informatics (LIPIcs)},
  ISBN =	{978-3-95977-082-8},
  ISSN =	{1868-8969},
  year =	{2018},
  volume =	{113},
  publisher =	{Schloss Dagstuhl--Leibniz-Zentrum fuer Informatik},
  address =	{Dagstuhl, Germany}
  }

@inproceedings{apm,
 author = {Heger, Christoph and van Hoorn, Andr{\'e} and Mann, Mario and Okanovi\'{c}, Du\v{s}an},
 title = {Application Performance Management: State of the Art and Challenges for the Future},
 booktitle = {Proceedings of the 8th ACM/SPEC on Intl. Conference on Performance Engineering},
 series = {ICPE '17},
 year = {2017},
 isbn = {978-1-4503-4404-3},
 location = {L'Aquila, Italy},
 pages = {429--432},
 numpages = {4},
 url = {http://doi.acm.org/10.1145/3030207.3053674},
 doi = {10.1145/3030207.3053674},
 acmid = {3053674},
 publisher = {ACM},
 address = {New York, NY, USA},
}

@Book{reussner2016a,
	author = {Reussner, Ralf H. and Becker, Steffen and Happe, Jens and Heinrich, Robert and Koziolek, Anne and Koziolek, Heiko and Kramer, Max and Krogmann, Klaus},
	title = {Modeling and Simulating Software Architectures -- The Palladio Approach},
	publisher = {MIT Press},
	month = oct,
	year = {2016},
	address = {Cambridge, MA},
	isbn = {9780262034760},
	pagetotal = {408},
	url = {http://mitpress.mit.edu/books/modeling-and-simulating-software-architectures},
	tags = {book}
}

@article{martens2010c,
	Author = {Anne Martens and Heiko Koziolek and Lutz Prechelt and Ralf Reussner},
	Doi = {10.1007/s10664-010-9142-8},
	Issn = {1382-3256},
	Journal = {Empirical Software Engineering},
	Number = {5},
	Pages = {587--622},
	Publisher = {Springer Netherlands},
	Title = {From monolithic to component-based performance evaluation of software architectures},
	Url = {http://dx.doi.org/10.1007/s10664-010-9142-8},
	Volume = {16},
	Year = {2011}, 
	tags = {peer-reviewed}}

@InProceedings{Mazkatli2018MiSE,
  author    = {Mazkatli, Manar and Burger, Erik and Quante, Jochen and Koziolek, Anne},
  title     = {Integrating semantically-related Legacy Models in Vitruvius},
  booktitle = {Proceedings of Modelling in Software Engineering co-located with the 40th International Conference on Software Engineering},
  year      = {2018},
  month     = may,
  doi       = {https://doi.org/10.1145/3193954.3193961},
  numpages  = {8},
  url       = {http://conferences.computer.org/icse-w/2018/pdfs/MiSE2018-5tgG6LGBeXFSudSNr7qTEN/5qDI4AQ7gU2ot7BLw0AvSq/6Sw5H8krW2oYLSaudkI0rk.pdf},
  tags={Vitruv}
}

@inproceedings{langhammer2016automated,
  title={Automated extraction of rich software models from limited system information},
  author={Langhammer, Michael and Shahbazian, Arman and Medvidovic, Nenad and Reussner, Ralf H},
  booktitle={2016 13th Working IEEE/IFIP Conference on Software Architecture (WICSA)},
   year={2016},
  organization={IEEE}
}

@InProceedings{Brunnert13,
author="Brunnert et al.",
title="Automatic Performance Model Generation for Java Enterprise Edition (EE) Applications",
booktitle="Computer Performance Engineering",
year="2013"
}

@inproceedings{becker2010reverse,
  title={Reverse engineering component models for quality predictions},
  author={Becker et al.},
  booktitle={2010 14th European Conference on Software Maintenance and Reengineering},
  year={2010},
  organization={IEEE}
}

@STRING{C = {IEEE Computer}}

@STRING{ELSEVIER = {Elsevier Science Publishers}}

@STRING{IS = {GI Informatik Spektrum}}

@STRING{OF = {Objekt Fokus}}

@STRING{SPE = {Software--Practice and Experience}}

@MISC{ATL,
  author = {{ATLAS Group}},
  title = {{Atlas Transformation Language (ATL) Homepage}},
  year = {2007},
  note = {Last retrieved 2008-01-06},
  key = {ATL},
  url = {http://www.eclipse.org/m2m/atl/}
}

@MISC{COM,
  author = {Microsoft Corp.},
  title = {The {COM} homepage},
  note = {http://www.microsoft.com/com/, last retrieved 2006-10-30},
  url = {http://www.microsoft.com/com/}
}

@MISC{emf,
  author = {{Eclipse Foundation}},
  title = {Eclipse Modeling Framework Homepage},
  howpublished = {\url{http://www.eclipse.org/modeling/emf/}},
  note = {last retrieved 2014-11-14},
  url = {http://www.eclipse.org/modeling/emf/}
}

@MISC{XTEXT,
  author = {{Eclipse Foundation}},
  title = {{xText} Website},
  note = {Last visited: 22nd of Febrary 2012},
  url = {http://www.xtext.org}
}

@incollection{heidenreich2010a,
year={2010},
isbn={978-3-642-12106-7},
booktitle={Software Language Engineering},
volume={5969},
series={Lecture Notes in Computer Science},
editor={van den Brand, Mark and Ga\v{s}evi\'{c}, Dragan and Gray, Jeff},
doi={10.1007/978-3-642-12107-4_25},
title={Closing the Gap between Modelling and Java},
url={http://dx.doi.org/10.1007/978-3-642-12107-4_25},
publisher={Springer Berlin Heidelberg},
author={Heidenreich, Florian and Johannes, Jendrik and Seifert, Mirko and Wende, Christian},
pages={374-383},
}

@MISC{DCOM,
  author = {{Microsoft Corporation}},
  title = {The {DCOM} homepage},
  year = {2007},
  note = {Last retrieved 2008-01-06},
  key = {DCOM},
  url = {http://www.microsoft.com/com/default.mspx}
}

@MISC{CORBA,
  author = {{Object Management Group (OMG)}},
  title = {{The {CORBA} homepage}},
  howpublished = {http://www.corba.org},
  key = {OMG}
}

@MISC{QVT,
  author = {{Object Management Group (OMG)}},
  title = {{Meta Object Facility (MOF) 2.0 Query/View/Transformation Specification
	(ptc/07-07-07)}},
  year = {2007},
  key = {QVT},
  publisher = {OMG},
  url = {http://www.omg.org/docs/ptc/07-07-07.pdf}
}

@MISC{CCM,
  author = {{Object Management Group (OMG)}},
  title = {{CORBA Component Model, v4.0 (formal/2006-04-01)}},
  year = {2006},
  key = {CCM},
  publisher = {OMG},
  url = {http://www.omg.org/technology/documents/formal/components.htm}
}

@MISC{MARTE,
  author = {{Object Management Group (OMG)}},
  title = {{UML Profile for Modeling and Analysis of Real-Time and Embedded
	systems (MARTE) RFP (realtime/05-02-06)}},
  year = {2006},
  key = {MARTE},
  publisher = {OMG},
  url = {http://www.omg.org/cgi-bin/doc?realtime/2005-2-6}
}

@MISC{MDA,
  author = {{Object Management Group (OMG)}},
  title = {{Model Driven Architecture - Specifications}},
  year = {2006},
  key = {MDA},
  publisher = {OMG},
  url = {http://www.omg.org/mda/specs.htm}
}

@MISC{OCL,
  author = {{Object Management Group (OMG)}},
  title = {{Object Constraint Language, v2.0 (formal/06-05-01)}},
  year = {2006},
  key = {OCL},
  publisher = {OMG},
  url = {http://www.omg.org/cgi-bin/doc?formal/2006-05-01}
}

@MISC{XMI,
  author = {{Object Management Group (OMG)}},
  title = {{Meta Object Facility (MOF) 2.0 XMI Mapping Specification, v2.1 (formal/05-09-01)}},
  year = {2006},
  key = {XMI},
  publisher = {OMG},
  url = {http://www.omg.org/cgi-bin/apps/doc?formal/05-09-01.pdf}
}

@MISC{oAW,
  author = {{openArchitectureWare (oAW)}},
  title = {{openArchitectureWare (oAW) Generator Framework}},
  year = {2007},
  note = {Last retrieved 2008-01-06},
  key = {openArchitectureWare},
  url = {http://www.openarchitectureware.org}
}

@MISC{JAVA,
  author = {{Sun Microsystems Corp.}},
  title = {{The {JAVA} homepage}},
  howpublished = {http://java.sun.com/},
  key = {JAVA}
}

@MISC{somox,
  title = {{SoMoX -- SOftware MOdel eXtractor}},
  note = {http://www.somox.org},
  url = {http://www.somox.org}
}

@MISC{EJB,
  title = {{Sun Microsystems Corp., The {Enterprise Java Beans} homepage}},
  year = {2007},
  note = {Last retrieved 2008-01-06},
  key = {EJB},
  url = {http://java.sun.com/products/ejb/}
}

@string{acm = {ACM, New York, NY, USA}}

@string{ieee = {IEEE}}

@string{jss = {Journal of Systems and Software}}

@string{springer = {Springer-Verlag Berlin Heidelberg}}

@article{becker2008a,
	Author = {Steffen Becker and Heiko Koziolek and Ralf Reussner},
	Doi = {10.1016/j.jss.2008.03.066},
	Journal = JSS,
	Pages = {3--22},
	Publisher = {Elsevier Science Inc.},
	Title = {{T}he {P}alladio component model for model-driven performance prediction},
	Url = {http://dx.doi.org/10.1016/j.jss.2008.03.066},
	Volume = {82},
	Year = {2009}}

@inproceedings{BrHuKo2011-ASE-AutomExtraction,
	Address = {Oread, Lawrence, Kansas},
	Author = {Fabian Brosig and Nikolaus Huber and Samuel Kounev},
	Booktitle = {26th IEEE/ACM Int. Conference On Automated Software Engineering (ASE 2011)},
	Month = nov,
	Title = {{A}utomated {E}xtraction of {A}rchitecture-{L}evel {P}erformance {M}odels of {D}istributed {C}omponent-{B}ased {S}ystems},
	Year = {2011}}

@phdthesis{burger2014diss,
	Address = {Karlsruhe, Germany},
	Author = {Erik Burger},
	Doi = {10.5445/KSP/1000043437},
	Editor = {Ralf Reussner},
	Isbn = {978-3-7315-0276-0},
	Issn = {1867-0067},
	Month = jul,
	Publisher = {KIT Scientific Publishing},
	School = {Karlsruhe Institute of Technology},
	Series = {The Karlsruhe Series on Software Design and Quality},
	Title = {{Flexible Views for View-based Model-driven Development}},
	Url = {http://digbib.ubka.uni-karlsruhe.de/volltexte/1000043437},
	Year = {2014},
	tags={Vitruv}}

@inproceedings{kramer2012b,
	
	Address = {Karlsruhe},
	Author = {Max E. Kramer and Zoya Durdik and Michael Hauck and J{\"o}rg Henss and Martin K{\"u}ster and Philipp Merkle and Andreas Rentschler},
	Booktitle = {Palladio Days 2012 Proceedings (appeared as technical report)},
	Pages = {7--15},
	Publisher = {KIT, Faculty of Informatics},
	Series = {Karlsruhe Reports in Informatics ; 2012,21},
	Tags = {workshop},
	Title = {{Extending the Palladio Component Model using Profiles and Stereotypes}},
	Url = {http://nbn-resolving.org/urn:nbn:de:swb:90-308043},
	Year = {2012}}

@inproceedings{kramer2013b,
	Acmid = {2489864},
	Address = {New York, NY, USA},
	Articleno = {5},
	Author = {Kramer, Max E. and Burger, Erik and Langhammer, Michael},
	Booktitle = {Proceedings of the 1st Workshop on View-Based, Aspect-Oriented and Orthographic Software Modelling},
	Doi = {10.1145/2489861.2489864},
	Isbn = {978-1-4503-2070-2},
	Location = {Montpellier, France},
	Numpages = {6},
	Pages = {5:1--5:6},
	Publisher = {ACM},
	Series = {VAO '13},
	Tags = {workshop, Vitruv},
	Title = {{View-Centric Engineering with Synchronized Heterogeneous Models}},
	Url = {http://doi.acm.org/10.1145/2489861.2489864},
	Year = {2013}}

@inproceedings{heinrich2014,
	Author = {Robert Heinrich and Eric Schmieders and Reiner Jung and Kiana Rostami and Andreas Metzger and Wilhelm Hasselbring and Ralf H. Reussner and Klaus Pohl},
	Booktitle = {Proceedings of the 9th Workshop on Models@run.time co-located with 17th Int. Conference on Model Driven Engineering Languages and Systems (MODELS 2014), Valencia, Spain, September 30, 2014.},
	Pages = {41--46},
	Title = {Integrating Run-time Observations and Design Component Models for Cloud System Analysis},
	Url = {http://ceur-ws.org/Vol-1270/mrt14_submission_8.pdf},
	Year = {2014},
	tags = {refereed}
	}

@inproceedings{jung2013model,
	Author = { Reiner Jung et al.},
	Booktitle = {Symposium on Software Performance},

	Publisher = {CEUR},
	Title = {Model-driven instrumentation with Kieker and Palladio to forecast dynamic applications},
	Url = {http://ceur-ws.org/Vol-1083/paper11.pdf},

	Year = {2013}}

@article{heinrich2016b,
	Author = {Heinrich, Robert},
	Journal = {ACM SIGMETRICS Performance Evaluation Review},
	Title = {Architectural Run-time Models for Performance and Privacy Analysis in Dynamic Cloud Applications},
	issue_date = {March 2016},
 volume = {43},
 number = {4},
 year = {2016},
 issn = {0163-5999},
 pages = {13--22},
 numpages = {10},
 url = {http://doi.acm.org/10.1145/2897356.2897359},
 doi = {10.1145/2897356.2897359},
 acmid = {2897359},
 publisher = {ACM},
 address = {New York, NY, USA},
}

@inproceedings{langhammer2015a,
	Author = {Langhammer, Michael and Krogmann, Klaus},
	Booktitle = {17. Workshop Software-Reengineering und-Evolution},
	Title = {A Co-evolution Approach for Source Code and Component-based Architecture Models},
	Url = {http://fg-sre.gi.de/fileadmin/gliederungen/fg-sre/wsre2015/WSRE2015-Proceeedings-preliminary.pdf#page=40},
	Volume = {4},
	Year = {2015}}

@InBook{heinrich2017a,
	author = {Robert Heinrich and Reiner Jung and Christian Zirkelbach and Wilhelm Hasselbring and Ralf Reussner},
	chapter = {An Architectural Model-Based Approach to Quality-aware DevOps in Cloud Applications},
	title = {Software Architecture for Big Data and the Cloud},
	publisher = {Elsevier},
	year = {2017},
	note = {to appear}
}

@inproceedings{WaStKoKo2017-QUDOS-PMXBuilder,
	author = {Walter, J\"{u}rgen and Stier, Christian and Koziolek, Heiko and Kounev, Samuel},
	title = {An Expandable Extraction Framework for Architectural Performance Models},
	booktitle = {Proceedings of the 8th ACM/SPEC on Int. Conference on Performance Engineering Companion},
	series = {ICPE '17 Companion},
	year = {2017},
	isbn = {978-1-4503-4899-7},
	location = {L'Aquila, Italy},
	pages = {165--170},
	numpages = {6},
	url = {http://doi.acm.org/10.1145/3053600.3053634},
	doi = {10.1145/3053600.3053634},
	acmid = {3053634},
	publisher = {ACM},
	address = {New York, NY, USA},
}

@phdthesis{langhammer2017a,
  author = {Michael Langhammer},
  title = {Automated Coevolution of Source Code and Software Architecture Models},
  school = {Karlsruhe Institute of Technology (KIT)},
  address = {Karlsruhe, Germany},
  year = {2017},
  pagetotal = {259 },
  doi = {10.5445/IR/1000069366},
  url = {http://nbn-resolving.org/urn:nbn:de:swb:90-693666},
  tags={Vitruv}
}
\clearpage
\end{document}